\documentclass[12pt,a4paper]{article}
\usepackage{amsmath,amssymb}
\usepackage[table]{xcolor}
  \usepackage{arydshln}
\usepackage{bm}
\usepackage{graphicx,here}
\usepackage{ascmac}
\usepackage{xcolor}
\usepackage{ctable,dcolumn}
\usepackage{slashbox}
\usepackage{multirow}
\usepackage{listliketab}
\usepackage[english]{babel}
\usepackage{setspace}
\def\vector#1{\mbox{\boldmath $#1$}}
\newcommand{\argmax}{\mathop{\rm arg~max}\limits}
\newcommand{\argmin}{\mathop{\rm arg~min}\limits}

\begin{document}
%
%
%
%
%
%

\begin{center}
\textbf{\LARGE Bayesian generalized fused lasso modeling via NEG distribution} 
\end{center}
\begin{center}
{\large Kaito Shimamura$^{1}$, \ Masao Ueki$^{2}$, \\
Shuichi Kawano$^{3}$ \ and \ Sadanori Konishi$^{4}$}
\end{center}

\begin{center}
\begin{minipage}{14cm}
{
\begin{center}
{\it {\footnotesize 

\vspace{1.2mm}


$^1$Department of Mathematics, Graduate School of Science and Engineering, \\ Chuo University, 
1-13-27, Kasuga, Bunkyo-ku, Tokyo 112-8551, Japan. \\

\vspace{1.2mm}

$^2$Biostatistics Center, Kurume University, \\
67, Asahimachi, Kurume-shi, Fukuoka 830-0011, Japan. \\

\vspace{1.2mm}

$^3$Graduate School of Information Systems,  The University of Electro-Communications, \\
1-5-1, Chofugaoka, Chofu-shi, Tokyo 182-8585, Japan. \\

\vspace{1.2mm}

$^4$Department of Mathematics, Faculty of Science and Engineering, Chuo University, \\
1-13-27, Kasuga, Bunkyo-ku, Tokyo 112-8551, Japan. \\
}}

\vspace{2mm}

ka1.618n@hotmail.co.jp \hspace{5mm} uekimrsd@nifty.com \\
skawano@ai.is.uec.ac.jp \hspace{5mm} konishi@math.chuo-u.ac.jp

\end{center}

\vspace{1mm} 

{\small {\bf Abstract:} 
The fused lasso penalizes a loss function by the $L_1$ norm for both the regression coefficients and their successive differences to encourage sparsity of both.
In this paper, we propose a Bayesian generalized fused lasso modeling based on a normal-exponential-gamma (NEG) prior distribution.
The NEG prior is assumed into the difference of successive regression coefficients.
The proposed method enables us to construct a more versatile sparse model than the ordinary fused lasso by using a flexible regularization term.
We also propose a sparse fused algorithm to produce exact sparse solutions.
Simulation studies and real data analyses show that the proposed method has superior performance to the ordinary fused lasso.

}

\vspace{3mm}



}
\end{minipage}
\end{center}
%
%
%
%
%
%
%
%
%
%
%
%
%
%
%
%
%
%
%
%
%
%
%

%
%
%
\section{Introduction}
%
%
%
%
%
%
%
%
%
%
%

A set of processes for selecting the best model using a model selection criterion such as Akaike information criterion (Akaike, 1973) or Bayesian information criterion (Schwarz, 1978) is effective for evaluating a regression model that has been estimated by maximum likelihood or the least-squares method (see, e.g., Konishi and Kitagawa (2008)). 
However, when analyzing high-dimensional data, the traditional method is not effective. 
Recently, new methods which can handle high-dimensional data such as regularization methods have been extensively studied.

In particular,
the $L_1$ norm regularization has attracted attention in various fields.
Lasso (Tibshirani, 1996) is a method of model estimation, which imposes the sum of absolute values ($L_1$ norms) of the regression coefficients as a constraint on the sum of squared errors. 
A distinctive feature of lasso is its capability for simultaneous model estimation and variable selection. 
In lasso, however, the $L_1$ norm constraint is non-differentiable at zero and no closed-form solution is available.
To address the difficulty,
various estimation algorithms for lasso have been developed such as the least angle regression (LARS) algorithm of  Efron \textit{et al.} (2004) and the coordinate descent algorithm of Friedman \textit{et al.} (2007).
Tibshirani \textit{et al.} (2005) proposed the fused lasso for the analysis of data where the predictor variables are in some sense ordered.
The fused lasso can be used for sparse modeling both for regression coefficients and for their successive differences.

Tibshirani (1996) demonstrated that the lasso estimates can be interpreted as a posterior mode estimation when the regression parameters have independent and identical Laplace (double-exponential) priors.
Park and Casella (2008) suggested Gibbs sampling for the lasso with a Laplace prior in a hierarchical model. 
Kyung \textit{et al.} (2010)  proposed a Bayesian fused lasso by interpreting the fused lasso in a Bayesian framework, assuming a product of the Laplace distribution in the prior of the regression coefficient vector.
However, methods which encourage sparsity between neighboring variables via the $L_1$ norm such as the fused lasso and Bayesian fused lasso may have a substantial bias in their estimators, because the ordinary methods impose a large penalty for differences between regression coefficients that belong to different groups.
As a result, the group difference is not contrasted, and then it may incur inaccuracy of prediction.

In this paper, we propose a Bayesian sparse fused lasso and a Bayesian sparse generalized fused lasso based on the normal-exponential-gamma (NEG) prior distribution. 
The NEG penalty allows construction of highly versatile sparse models, because it has spike at zero and more extreme flatness in its tail than does the lasso penalty (Griffin and Brown, 2005; Hoggart {\it et al.}, 2008).
Using a NEG prior to the difference of successive regression coefficients, our Bayesian sparse modeling can yield clearly different estimates for parameters in different groups and improves prediction accuracy.

The rest of this paper is organized as follows. 
Section 2 describes the $L_1$ norm regularization.
In Section 3, we describe the Bayesian sparse modeling which formulates the sparse estimation in a Bayesian framework.
In Section 4, we propose a Bayesian sparse modeling having higher versatility than the
fused lasso by using the NEG distribution. 
Monte Carlo simulations and real data analysis are
conducted to examine the performance of our proposed procedure and to compare it with
existing methods in Section 5. 
Concluding remarks are given in Section 6.

%
%
%
%
%
%
%
%
%
%
%
%
%
%
%
\section{$L_1$ norm regularization}
%
%
%

In this section, 
we describe the $L_1$ norm regularization,
where the sum of absolute values of regression coefficients is imposed in a penalty term.
In particular, we describe the lasso, fused lasso, and generalized fused lasso.

%
%
%
%
%
%
\subsection{Regularized likelihood method}
%
%
%
%
%
%

Suppose that we have observed data $\{(y_i,\vector{x}_i); i=1,2,\dots,n\}$ for response variable $y$ and $p$-dimensional predictor variables $\vector{x}=(x_1,x_2,\dots,x_p)^T$. 
Without loss of generality, the response is centered around the mean and the predictors are standardized:
\[
\sum_{i=1}^n y_i =0, \quad \sum_{i=1}^n x_{ij} = 0, \quad \sum_{i=1}^n x_{ij}^2 = n \quad (j=1, 2, \dots, p).
\]
We consider the following linear regression model without the intercept:
\begin{eqnarray}
\vector{y} = X \vector{\beta} + \vector{\epsilon}, \label{model}
\end{eqnarray}
where $\vector{y} = (y_1,y_2,\dots,y_n)^T$ is the $n$-dimensional vector of observed values for the response variable, $X=({\bm x}_1, \ldots, {\bm x}_n)^T$ is the $n\times p$ design matrix, $\vector{\beta}$ is the $p$-dimensional regression coefficient vector, and $\vector{\epsilon}$ is the $n$-dimensional error vector distributed as $\mbox{N}_n \left(\vector{0}_n, \sigma^2I_n\right)$. 
Since the error vector $\vector{\epsilon}$ is distributed as multivariate normal distribution with mean $\vector{0}_n$ and variance-covariance matrix $\sigma^2I_n$, the likelihood function is given by
\begin{eqnarray}
f(\vector{y}|X;\vector{\beta},\sigma^2)
&=&
\prod_{i=1}^n f(y_i|\vector{x}_i;\vector{\beta},\sigma^2) 
,
\end{eqnarray}
where
\[
f(y_i|\vector{x}_i;\vector{\beta},\sigma^2)
=
\left(2\pi\sigma^2\right)^{-1/2}
\exp\left\{
-\frac{(y_i-\vector{x}_i^T\vector{\beta})^2}{2\sigma^2}
\right\}.
\]
Hereafter, we denote the probability density function $f(y_i|\vector{x}_i;\vector{\beta},\sigma^2)$ as $f(y_i|\vector{\beta},\sigma^2)$ for simplicity.
%

A regularization method 
imposes a constraint condition for $\vector{\beta}$ 
with a penalty function $P(\vector{\beta})\ (>0)$ 
on the maximization of the loss function such as a log-likelihood function $\log f(\vector{y}|\vector{\beta},\sigma^2)$.
We consider the following constrained optimization problem:
\begin{eqnarray}
\max_{\vector{\beta}}
\sum_{i=1}^n \log f(y_i|\vector{\beta}, \sigma^2),\ \ 
\mbox{subject to}\ P(\vector{\beta}) \leq t
,
\label{optim_prob}
\end{eqnarray}
where $t \ (\geq 0)$ is a constant.
The above optimization problem is equivalent to the maximization of the following objective function,
\begin{eqnarray}
\sum_{i=1}^n \log f(y_i|\vector{\beta},s \sigma^2) - p_\gamma(\vector{\beta})
,
\label{eqqq}
\end{eqnarray}
where $p_\gamma(\vector{\beta}) \ (>0)$ is a penalty function corresponding to the constraint $P(\vector{\beta}) \leq t$ and $\gamma\ (>0)$ is a tuning parameter to control the degree of penalties, called the regularization parameter.
When $p_\gamma(\vector{\beta})=\gamma\|\vector{\beta}\|_2^2$, the optimization problem ($\ref{eqqq}$) reduces to the ridge regression problem proposed by Hoerl and Kennard (1970).
The ridge regression improves the prediction performance, but it cannot produce zero values for regression coefficients.
%

%
%
%
%
%
%
%
%
%
\subsection{Lasso}
%
%
%
%
%
%
%
%
%

When $p_\gamma(\vector{\beta})=\gamma \sum_{j=1}^p |\beta_j|$, the optimization problem ($\ref{eqqq}$) reduces to the lasso problem by Tibshirani (1996):
\begin{eqnarray}
\hat{\vector{\beta}}=
\argmax_{\vector{\beta}}
\left\{
\log f(\vector{y}|\vector{\beta},\sigma^2)
-\gamma \sum_{j=1}^p |\beta_j|
\right\}
.
\label{lasso_opt}
\end{eqnarray}
In contrast to the shrinkage of regression coefficients toward zero that occurs in ridge regression, the lasso results in exactly zero estimates for some of the coefficients. 
The regularization parameter $\gamma$ controls the overall model sparsity (that is, the model with exactly zero values for the coefficients) and shrinkage of the regression coefficients.
A larger value of the regularization parameter produces sparser models.
%
%
%
%
%
%
%
%
%
%
%
%
%
%
%
%
%
\subsection{Fused lasso}
%
%

Tibshirani {\it et al}. (2005) proposed the fused lasso for the sake of analyzing data whose predictor variables are in some sense ordered.
The regularization procedure gives estimates by
\[
\hat{\vector{\beta}}
=
\argmax_{\vector{\beta}}\left\{
\log f(\vector{y}|\vector{\beta},\sigma^2)
-\lambda_1\sum_{j=1}^p |\beta_j|
-\lambda_2\sum_{j=2}^p |\beta_j-\beta_{j-1}|
\right\}
,
\]
where $\lambda_1\ (>0)$ and $\lambda_2\ (>0)$ are regularization parameters. 
The $\lambda_1$ controls the degree of sparsity and $\lambda_2$ controls the degree of smoothing between successive differences.
If $\lambda_2=0$, the fused lasso reduces to the lasso.
In recent years, 
the fused lasso has become the focus of increasing interest as a useful technique in genomic data analysis, 
image processing, and many other field (see, e.g., Friedman {\it et al.} (2007), Tibshirani and Wang (2008)). 
The upper left panel of Figure \ref{FlassoNEG_penal} shows the penalty 
\begin{eqnarray}
p_{\lambda_2}(\beta_j)=
\lambda_2\Big(
|\beta_j-\beta_{j-1}|+|\beta_{j+1}-\beta_j|
\Big)
\label{fused_penalty_f}
\end{eqnarray}
as a function of $\beta_j$, while we fix both $\beta_{j-1}$ and $\beta_{j+1}$.

A general form of the generalized fused lasso is given by
\[
\hat{\vector{\beta}}
=
\argmax_{\vector{\beta}}\left\{
\log f(\vector{y}|\vector{\beta},\sigma^2)
-\lambda_1\sum_{j=1}^p |\beta_j|
-\lambda_2\sum_{(j,k)\in E} |\beta_j-\beta_k|
\right\}
,
\]
where $E \subset \{(j,k);j,k=1,\dots,p\}$. 
It is important to determine the set $E$ according to the subject of the analysis.
Examples of the generalized fused lasso include hexagonal operator for regression with
shrinkage and equality selection (HORSES; Jang {\it et al}., 2013), 
which is a regularization method that maximizes the objective function
\[
\log f(\vector{y}|\vector{\beta},\sigma^2)
-\lambda_1\sum_{j=1}^p |\beta_j|
-\lambda_2\sum_{j>k} |\beta_j-\beta_k|
.
\]
In HORSES, all combinations between two regression coefficients are used as a penalty.
Although in the fused lasso, the predictors must be in some sense ordered, 
HORSES, on the other hand, does not require that condition.

One of useful applications of the fused lasso is the fused lasso signal approximator (FLSA; Friedman {\it et al.}, 2007). 
The FLSA solves the optimization problem
\begin{eqnarray}
\min_{\beta_1,\dots,\beta_n}\left\{
\sum_{i=1}^n (y_i-\beta_i)^2
+\lambda_1\sum_{i=1}^n|\beta_i|
+\lambda_2\sum_{i=2}^n|\beta_i-\beta_{i-1}|
\right\}.
\end{eqnarray}
The FLSA corresponds to the case where $n=p$ and $X=I_n$ in the ordinary fused lasso.
Tibshirani and Wang (2008) applied the FLSA to the analysis of comparative genomic hybridization (CGH) data.
\section{Bayesian sparse modeling via Gibbs sampling}
In this section, we describe the Bayesian lasso which formulates the lasso in a Bayesian framework. 
We consider the Bayesian sparse estimation with an NEG distribution as the prior distribution instead of the Laplace prior distribution. 
In addition, the Bayesian fused lasso is described to formulate the fused lasso in a Bayesian framework.

%
%
%
%
%
%
%
%
%
%
%
%
%
%
%
%
%
%
%
%
%

%
%
%
%
%
%
\subsection{Bayesian lasso}
%
%
%
The posterior distribution of coefficient vector $\vector{\beta}$ is given by
\[
\pi(\vector{\beta}|\vector{y}) \propto
f(\vector{y}|\vector{\beta},\sigma^2)\pi(\vector{\beta}|\sigma^2)\pi(\sigma^2)
.
\]
The coefficient vector $\vector{\beta}$ is estimated by the posterior mode for given data $\vector{y}$.
Park and Casella (2008) proposed to assume the Laplace prior on the coefficient vector $\vector{\beta}$:
\begin{eqnarray}
\pi(\vector{\beta}|\sigma^2)
=
\prod_{j=1}^p
\frac{\lambda}{2\sqrt{\sigma^2}}
\exp
\left(
-\frac{\lambda}{\sqrt{\sigma^2}}|\beta_j|
\right)
\label{Lap_pri}
\end{eqnarray}
and the non-informative scale-invariant prior $\pi(\sigma^2)=1/\sigma^2$ or inverse-gamma prior $\pi(\sigma^2)=\mbox{IG}(\nu_0/2,\eta_0/2)$ on $\sigma^2$, where $\nu_0 \ (>0)$ is a shape parameter and $\eta_0 \ (>0)$ is a scale parameter. 
An inverse-gamma probability density function is given by
\[
\mbox{IG}(x|\nu, \eta) = \frac{\eta^\nu}{\Gamma(\nu)}x^{-(\nu+1)}\exp\left( -\frac{\eta}{x} \right),
\]
where $\Gamma(\cdot)$ is the gamma function. 
The hyper-parameter $\lambda$ in ($\ref{Lap_pri}$) plays the same role as that of regularization parameter $\gamma$ in (\ref{lasso_opt}). 
It controls the degree of sparsity of the coefficients estimated. 
In other words, 
the larger values of hyper-parameter $\lambda$ get, the more numbers of zero regression coefficients increase. 
The smaller values of $\lambda$ get, the less numbers of zero regression coefficients increase.

The Laplace distribution is represented by a scale mixture of normals (Andrews and Mallows, 1974):
\begin{eqnarray*}
\frac{\lambda}{2\sqrt{\sigma^2}}\exp\left(-\frac{\lambda}{\sqrt{\sigma^2}}|\beta|\right)
=
\int_0^\infty
\frac{1}{\sqrt{2\pi\sigma^2\tau^2}}\exp\left(-\frac{\beta^2}{2 \sigma^2 \tau^2}\right)
\frac{\lambda^2}{2}\exp\left(-\frac{\lambda^2}{2}\tau^2\right)
d\tau^2
.
\end{eqnarray*}
From this relationship, Park and Casella (2008) assumed the following priors:
\begin{eqnarray*}
\pi(\vector{\beta}|\sigma^2, \tau_1^2, \tau_2^2, \dots, \tau_p^2) 
&=& 
\prod_{j=1}^p \frac{1}{\sqrt{2\pi \sigma^2 \tau_j^2}} 
\exp \left( -\frac{\beta_j^2}{2\sigma^2\tau_j^2} \right)
, 
\\
\pi(\tau_1^2, \tau_2^2, \dots,\tau_p^2)
&=&
\prod_{j=1}^p \frac{\lambda^2}{2} \exp \left( -\frac{\lambda^2}{2}\tau_j^2 \right)
.
\end{eqnarray*}
As a result, it enables us to carry out Bayesian estimation by Gibbs sampling.
Assuming an inverse-gamma prior $\mbox{IG}(\nu_0/2,\eta_0/2)$ on $\sigma^2$: 
\[
\pi(\sigma^2)
=
\frac{(\eta_0/2)^{\nu_0/2}}{\Gamma(\nu_0/2)}
(\sigma^2)^{-(\nu_0/2+1)}
\exp\left(
-\frac{\eta_0/2}{\sigma^2}
\right)
,
\]
the full-conditional posteriors on $\vector{\beta}, \sigma^2, \tau_1^2, \tau_2^2, \dots,\tau_p^2$ are given by
\begin{eqnarray*}
\vector{\beta}|\vector{y}, X, \sigma^2, \tau_1^2, \tau_2^2, \dots, \tau_p^2 
&\sim&
\mbox{N}_p(A^{-1}X^T\vector{y},\ \sigma^2A^{-1}) 
,
\\
&&
A=X^TX +  D_r^{-1} ,\quad  D_r = \mbox{diag}(\tau_1^2,\ \tau_2^2,\ \dots,\ \tau_p^2) 
,
\\
\sigma^2|\vector{y}, X, \vector{\beta}, \tau_1^2, \tau_2^2, \dots, \tau_p^2
&\sim&
\mbox{IG}\left(\frac{\nu_1}{2},\ \frac{\eta_1}{2}\right) 
,
\\
&&
\nu_1 = n+p+\nu_0,\quad \eta_1 = \| \vector{y} - X\vector{\beta} \|^2_2 +
 \vector{\beta}^TD_r^{-1}\vector{\beta}+\eta_0 ,\\
\left.\  \frac{1}{\tau_j^2} \right| \beta_j, \sigma^2, \lambda 
&\sim&
\mbox{IGauss} (\mu',\ \lambda') 
,
\\
&&
\mu' = \sqrt{\frac{\lambda^2\sigma^2}{\beta_j^2}},\quad \lambda'=\lambda^2,\quad j=1, 2, \dots, p
,
\end{eqnarray*}
where $\mbox{IGauss}(\mu, \lambda)$ denotes the inverse-Gaussian distribution with a density function
\[
\sqrt{\frac{\lambda}{2\pi}}x^{-3/2}\exp\left\{ -\frac{\lambda(x-\mu)^2}{2\mu^2x} \right\}
\quad (x>0).
\]

%
%
%
%
%
%
%
%
%
%
%
%
\subsection{Bayesian fused lasso}
%
%
%

Kyung {\it et al}. (2010) proposed the Bayesian fused lasso by interpreting the fused lasso in a Bayesian framework. 
In the Bayesian fused lasso, the prior distribution of the regression coefficients $\vector{\beta}$ is defined as follows:
\begin{eqnarray*}
\pi(\vector{\beta}|\sigma^2)
\propto
(\sigma^2)^{-\frac{2p-1}{2}}
\exp\left(
-\frac{\lambda_1}{\sigma}\sum_{j=1}^p|\beta_j|
-\frac{\lambda_2}{\sigma}\sum_{j=2}^p|\beta_j-\beta_{j-1}|
\right).
\end{eqnarray*}

This can be expressed as a hierarchical representation of the Laplace distribution,
\begin{eqnarray*}
\pi(\vector{\beta}|\sigma^2)
&\propto&
(\sigma^2)^{-\frac{2p-1}{2}}
\prod_{j=1}^p \int
\frac{1}{\sqrt{2\pi \tau_j^2}}\exp\left( -\frac{\beta_j^2}{2\sigma^2\tau_j^2} \right)
\frac{\lambda_1^2}{2}\exp\left( -\frac{\lambda_1^2}{2}\tau_j^2 \right)
d\tau_j^2 
\\
&& \qquad \times
\prod_{j=2}^p \int
\frac{1}{\sqrt{2\pi\widetilde{\tau}_j^2}}\exp\left\{ -\frac{(\beta_j-\beta_{j-1})^2}{2\sigma^2\widetilde{\tau}_j^2} \right\}
\frac{\lambda_2^2}{2}\exp\left( -\frac{\lambda_2^2}{2}\widetilde{\tau}_j^2 \right)
d\widetilde{\tau}_j^2 
\\
&\propto&
\int\int(\sigma^2)^{-\frac{2p-1}{2}}
\prod_{j=1}^p(\tau_j^2)^{-\frac{1}{2}} \prod_{j=2}^p(\widetilde{\tau}_j^2)^{-\frac{1}{2}}
\exp\left(-\frac{1}{2\sigma^2}\vector{\beta}^T\Sigma_{\vector{\beta}}^{-1}\vector{\beta}\right)
\\
&& \qquad \times
\prod_{j=1}^p\pi(\tau_j^2)\prod_{j=2}^p\pi(\widetilde{\tau}_j^2)
\prod_{j=1}^p d\tau_j^2 \prod_{j=2}^p d\widetilde{\tau}_j^2
,
\end{eqnarray*}
where
\begin{eqnarray}
\Sigma_{\vector{\beta}}^{-1}
=
\left\{
\begin{array}{cccccc}
\frac{1}{\tau_1^2}+\frac{1}{\widetilde{\tau}_2^2} & -\frac{1}{\widetilde{\tau}_2^2} & 0 & \cdots & 0 & 0 \\
 -\frac{1}{\widetilde{\tau}_2^2} &  \frac{1}{\tau_2^2}+\frac{1}{\widetilde{\tau}_2^2}+\frac{1}{\widetilde{\tau}_3^2} & -\frac{1}{\widetilde{\tau}_3^2} & \cdots & 0 & 0 \\
0 & -\frac{1}{\widetilde{\tau}_3^2} & \frac{1}{\tau_3^2}+\frac{1}{\widetilde{\tau}_3^2}+\frac{1}{\widetilde{\tau}_4^2} & \cdots & 0 & 0 \\
\vdots & \vdots & \vdots & \ddots & \vdots & \vdots \\
0 & 0 & 0 & \cdots & \frac{1}{\tau_{p-1}^2}+\frac{1}{\widetilde{\tau}_{p-1}^2}+\frac{1}{\widetilde{\tau}_{p}^2} & -\frac{1}{\widetilde{\tau}_{p}^2} \\
0 & 0 & 0 & \cdots & -\frac{1}{\widetilde{\tau}_{p}^2} & \frac{1}{\tau_p^2}+\frac{1}{\widetilde{\tau}_{p}^2}
\end{array}
\right\}.
\label{Sigmabeta}
\end{eqnarray}
This formulation enables us to implement Gibbs sampler for $\vector{\beta}, \sigma^2, \tau_1^2, \tau_2^2, \dots,\tau_{p}^2$ and $\widetilde{\tau}_2^2,\widetilde{\tau}_3^2,\dots,\widetilde{\tau}_{p}^2$. 
The full-conditional distribution is then given by
\begin{eqnarray*}
\vector{\beta}|\vector{y},X,\sigma^2,\tau_1^2,\tau_2^2,\dots,\tau_p^2,\widetilde{\tau}_2^2,\widetilde{\tau}_3^2,\dots,\widetilde{\tau}_{p}^2
&\sim&
\mbox{N}_p\left( (X^TX+\Sigma^{-1}_{\vector{\beta}})^{-1}X^T\vector{y},\ \sigma^2(X^TX+\Sigma^{-1}_{\vector{\beta}})^{-1} \right)
,
\\
\sigma^2|\vector{y},X,\vector{\beta},\tau_1^2,\tau_2^2,\dots,\tau_p^2,\widetilde{\tau}_2^2,\widetilde{\tau}_3^2,\dots,\widetilde{\tau}_{p}^2
&\sim&
\mbox{IG}\left( \nu_1/2 ,\ \eta_1/2 \right)
,
\\
&&
\nu_1=n+2p-1+\nu_0
,
\\
&&
\eta_1=
(\vector{y}-X\vector{\beta})^T(\vector{y}-X\vector{\beta})+\vector{\beta}^T\Sigma^{-1}_{\vector{\beta}}\vector{\beta}+\eta_0
,
\\
\frac{1}{\tau_j^2}|\beta_j,\sigma^2,\lambda_1
&\sim&
\mbox{IGauss}\left( \sqrt{\frac{\lambda_1^2\sigma^2}{\beta_j^2}},\ \lambda_1^2 \right)
,
\\
\frac{1}{\widetilde{\tau}_j^2}|\beta_j,\beta_{j-1},\sigma^2,\lambda_2
&\sim&
\mbox{IGauss}\left( \sqrt{\frac{\lambda_2^2\sigma^2}{(\beta_j-\beta_{j-1})^2}},\ \lambda_2^2 \right)
,
\end{eqnarray*}
where an inverse-gamma prior distribution $\mbox{IG}(\nu_0/2,\eta_0/2)$ is assumed for $\sigma^2$. 

%
%
%
%
%
%
\subsection{Lasso-type Bayesian sparse regression via NEG prior}
%
%
%

Griffin and Brown (2005) proposed an NEG distribution as a prior distribution for the regression coefficients $\vector{\beta}$ which leads to more flexible with respect to sparsity than a Laplace distribution.
The NEG density function is given by
\begin{eqnarray}
\mbox{NEG}(\beta_j|\lambda,\gamma)
=
\kappa\exp\left(\frac{\beta_j^2}{4\gamma^2}\right)
D_{-2\lambda-1}\left(\frac{|\beta_j|}{\gamma}\right)
,
\label{NEG_dist}
\end{eqnarray}
where 
$\kappa = \displaystyle (2^\lambda\lambda)/(\gamma\sqrt{\pi})\Gamma(\lambda+1/2)$ is a normalization constant and
$D_{-2\lambda-1}$ is a parabolic cylinder function. 
The parabolic cylinder function is a solution of the second-order linear ordinary differential equation
\[
\frac{d^2w}{dz^2}
-\left(
\frac{z^2}{4}-\frac{1}{2}-a
\right)w=0
,
\]
and its integral representation is given by
\begin{eqnarray*}
D_{-2\lambda-1}\left(\frac{|\beta|}{\gamma}\right)
=
\frac{1}{\Gamma(2\lambda+1)}\exp\left(-\frac{\beta^2}{4\gamma^2}\right)
\int_0^\infty
w^{2\lambda}
\exp\left(
-\frac{1}{2}w^2-\frac{|\beta|}{\gamma}w 
\right)
dw
.
\end{eqnarray*}

Then, NEG density function can be expressed as a hierarchical representation
\begin{eqnarray*}
&&
\mbox{NEG}\left(\beta_j|\lambda,\gamma\right)
\\
&=&
\int\int
\frac{1}{\sqrt{2\pi\tau_j^2}}\exp\left(-\frac{\beta_j^2}{2\tau_j^2}\right)
\psi_j \exp\left(-\psi_j\tau_j^2\right)
\frac{(\gamma^2)^\lambda}{\Gamma(\lambda)}\psi_j^{\lambda-1}\exp\left(-\gamma^2\psi_j\right)
d\tau_j^2d\psi_j
\\
&=&
\int\int
\mbox{N}(\beta_j|0,\tau_j^2)
\mbox{EXP}(\tau_j^2|\psi_j)
\mbox{Ga}(\psi_j|\lambda,\gamma^2)
d\tau_j^2d\psi_j.
\end{eqnarray*}

The lasso-type Bayesian sparse estimation via an NEG distribution (Rockova and Lesaffre, 2014) assumes the following the NEG distribution instead of the Laplace distribution as a prior distribution for the regression coefficients $\vector{\beta}$,
\begin{eqnarray*}
\pi(\vector{\beta}|\sigma^2)
&=&
\prod_{j=1}^p
\frac{1}{\sqrt{\sigma^2}}\mbox{NEG}\left(\frac{\beta_j}{\sqrt{\sigma^2}}\Big|\lambda,\gamma\right).
\end{eqnarray*}
By assuming the above prior distribution, it is possible to guarantee a unimodal posterior distribution (Rockova and Lesaffre, 2014) and perform Bayesian estimation of the regression coefficient vector by Gibbs sampling in the same way as the Bayesian lasso.
The full-conditional distributions of 
$\vector{\beta}, \sigma^2, 1/\tau_j^2$ and $\psi_j\ (j=1,2, \dots p)$ are given by
\begin{eqnarray*}
\vector{\beta}|\vector{y},X,\sigma^2,\tau_1^2,\tau_2^2,\dots,\tau_p^2
&\sim&
\mbox{N}_p(A^{-1}X^T\vector{y},\sigma^2A^{-1})
,
\\
&& 
A=X^TX+D_r^{-1},\quad
D_r = \mbox{diag} (\tau_1^2,\dots,\tau_p^2)
,
\\
\sigma^2|\vector{y},X,\vector{\beta},\tau_1^2,\tau_2^2,\dots,\tau_p^2
&\sim&
\mbox{IG}(\nu_1/2,\eta_1/2)
,
\\
&&
\nu_1 = n+p+\nu_0,\quad 
\eta_1=\|\vector{y}-X\vector{\beta}\|_2^2+\vector{\beta}^TD^{-1}_r\vector{\beta}+\eta_0
,
\\
\frac{1}{\tau^2_j}|\beta_j,\sigma^2,\psi_j
&\sim&
\mbox{IGauss}(\mu',\lambda'),\ j=1,2,\dots,p
,
\\
&&
\mu'=\sqrt{\frac{2\psi_j\sigma^2}{\beta_j^2} },\quad
\lambda'=2\psi_j
,
\\
\psi_j|\tau_j^2,\lambda,\gamma
&\sim&
\mbox{Ga}(\lambda+1,\tau_j^2+\gamma^2),\ j=1,2,\dots,p
.
\end{eqnarray*}

\begin{figure}[t]
\vspace{-15mm}
 \begin{minipage}{2\hsize}
   \includegraphics[bb=0 0 716 715 ,width=100mm]{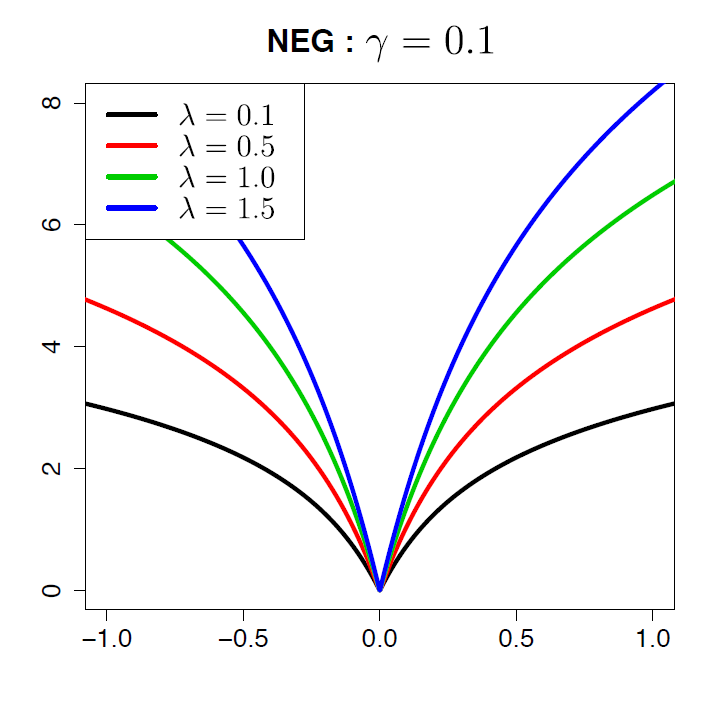}
\hspace{-15mm}
   \includegraphics[bb=0 0 716 715 ,width=100mm]{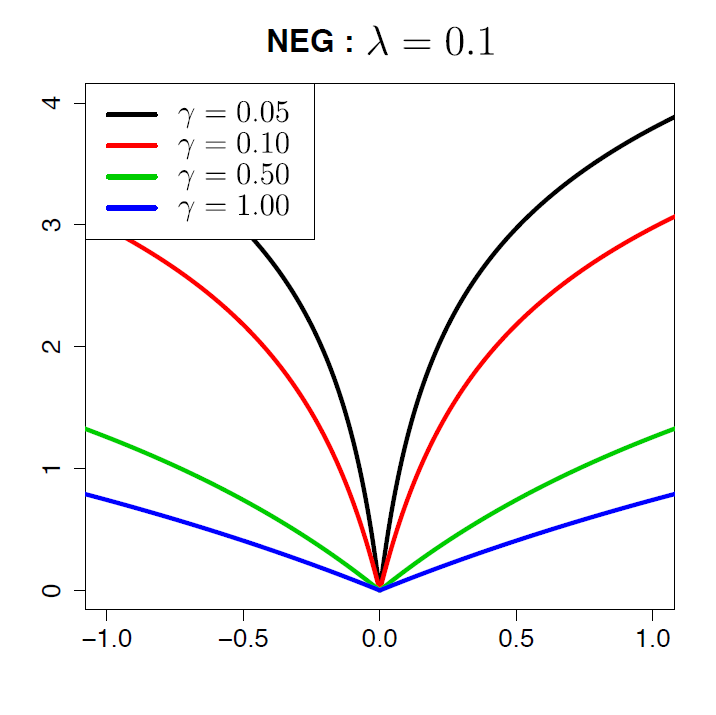}
 \end{minipage}\\
 \caption{
The NEG penalty function in Equation ($\ref{NEG_penalty_f}$),
$p_{\lambda,\gamma}(\beta)=\log\mbox{NEG}(\beta|\lambda,\gamma)+C$. 
The left panel shows functions under varying $\lambda$ at $\gamma=0.1$, while the right panel shows those under varying $\gamma$ at $\lambda=0.1$.
}
\label{fig:anthi02}
\end{figure}

\begin{figure}[htbp]
  \centering
 \begin{minipage}{0.32\columnwidth}
  \centering
   \includegraphics[bb=0 0 717 715 ,width=55mm]{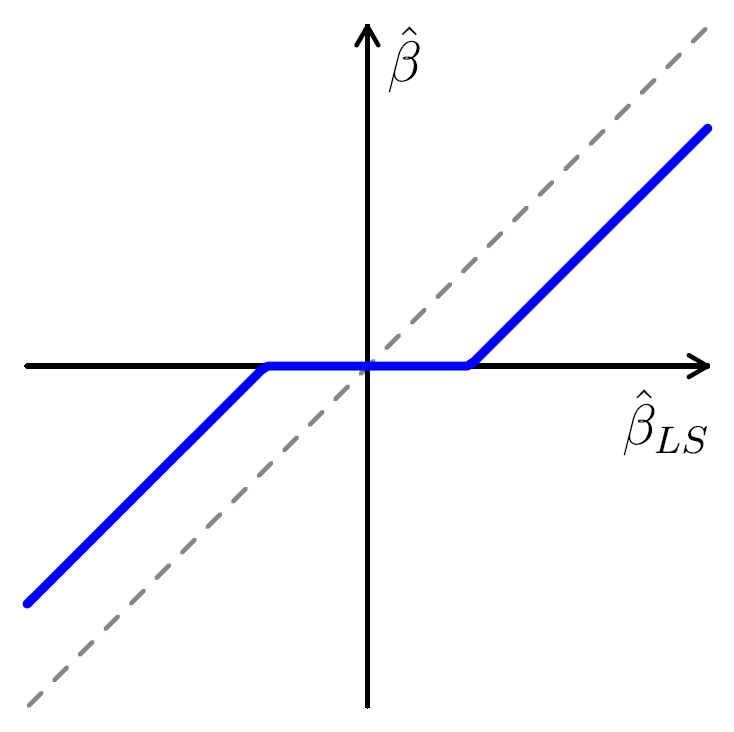}
 \end{minipage}
 \begin{minipage}{0.32\columnwidth}
  \centering
   \includegraphics[bb=0 0 711 710 ,width=55mm]{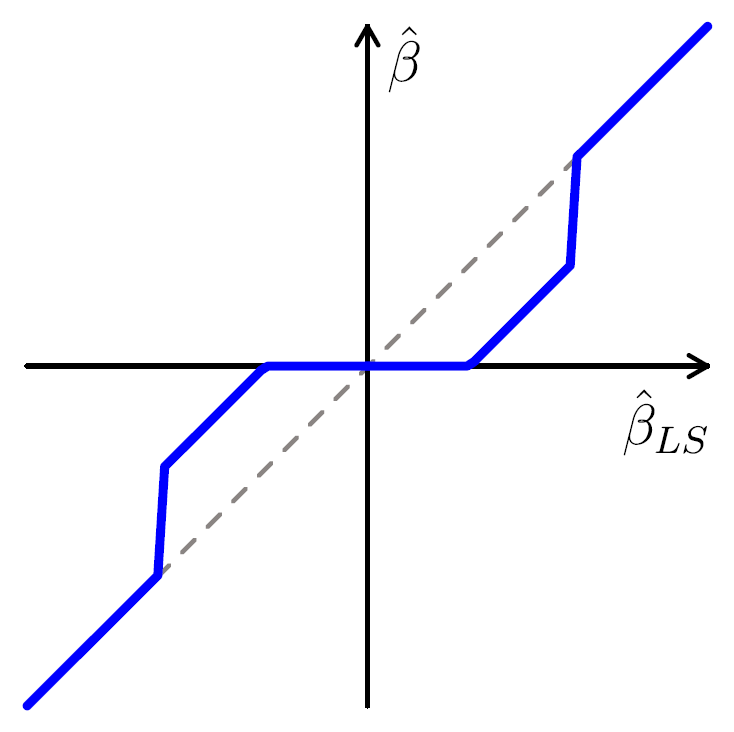}
 \end{minipage}
 \begin{minipage}{0.32\columnwidth}
  \centering
   \includegraphics[bb=0 0 717 715 ,width=55mm]{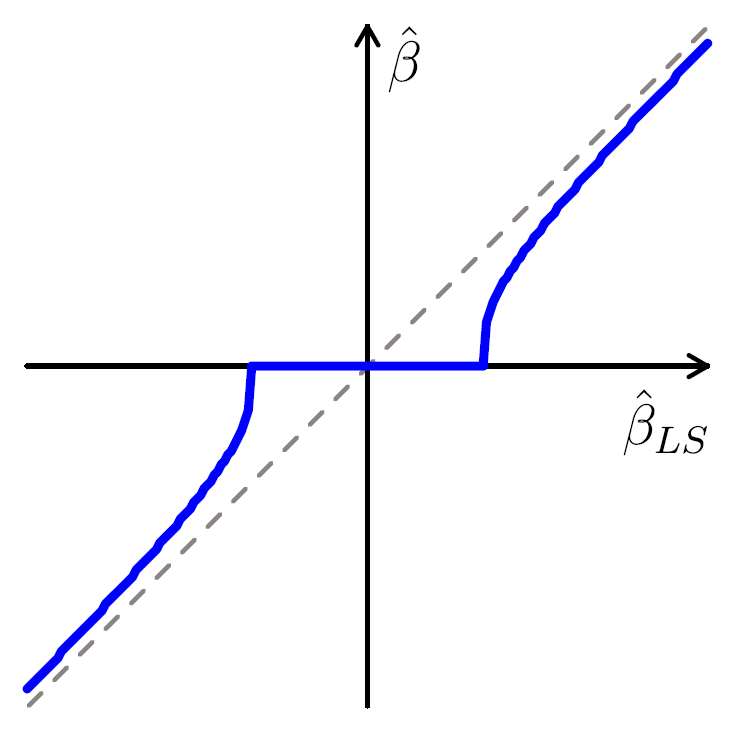}
 \end{minipage}
  \caption{
The relationship between the least-squares estimator and shrinkage estimator for lasso (left panel), SCAD (middle panel) and NEG (right panel). The dotted lines are the least-squares estimator $\hat{\beta}_{LS}$, while the solid lines are shrinkage estimators.
}
  \label{Est_bias}
\end{figure}

The NEG distribution can maintain flat tails with a large preponderance of the density around zero, 
making the resulting estimator more clear-cut.
As both $\lambda$ and $\gamma$ increase such that $\xi=\sqrt{2\lambda}/\gamma$ remains a constant, the NEG distribution converges to the Laplace distribution with a parameter $\xi$.
The NEG distribution is differentiable everywhere except at the point $0$. 
First and second derivatives of the NEG density function at $\beta\neq 0$ are respectively given by
\begin{eqnarray}
\displaystyle
\frac{\partial}{\partial \beta} \mbox{NEG}(\beta)
&=&
-\kappa
\frac{2(\lambda+1/2)\mbox{sign}(\beta)}{\gamma}
\exp\left(\frac{\beta^2}{4\gamma^2}\right)
D_{-(2\lambda+2)}\left(\frac{|\beta|}{\gamma}\right)
\label{NEG_da}
,
\\
\displaystyle
\frac{\partial^2}{\partial\beta^2} \mbox{NEG}(\beta)
&=&
\kappa\frac{4(\lambda+1/2)(\lambda+1)}{\gamma^2}
\exp\left(
\frac{\beta^2}{4\gamma^2}
\right)
D_{-(2\lambda+3)}\left(\frac{|\beta|}{\gamma}\right)
.
\end{eqnarray}
Figure \ref{fig:anthi02} shows the NEG penalty function
\begin{eqnarray}
p_{\lambda,\gamma}(\beta)=
\log\mbox{NEG}(\beta|\lambda,\gamma)+
C
,
\label{NEG_penalty_f}
\end{eqnarray}
when the regularization parameters are varied,
where $C$ is a constant such that $p_{\lambda,\gamma}(\beta)$ takes zero value at $\widetilde{\beta}=\argmin p_{\lambda,\gamma}(\beta)$.
The regularization parameters $\lambda$ and $\gamma$ affect the degree of sparsity of the solution: either a larger value of $\lambda$ or a smaller value of $\gamma$ produces sparser results. 
Setting an appropriate value of the regularization parameters is an important problem.
Rockova and Lesaffre (2014) summarized the properties of the NEG distribution.
The most remarkable property is
\begin{eqnarray*}
\displaystyle \frac{\partial}{\partial \beta} \log \mbox{NEG} \left(\beta\big|\lambda,\gamma\right)=\mathcal{O}\left(\frac{1}{|\beta|}\right) \quad {\rm as} \ \  |\beta|\rightarrow \infty,
\end{eqnarray*}
which implies that the regression estimator is less biased for large $|\beta|$.
%
%
%
%
%
%
%
%
%
%
%
%
%
%
%
%
%
%
%
The lasso estimator varies continuously, but is highly biased because of the strong constraint imposed on nonzero estimates. 
It will be more clear by considering the univariate least-squares problem,
%
%
%
%
%
%
%
\begin{eqnarray}
\hat{\beta}
=
\argmin_{\beta}\left\{
\frac{1}{2}(\hat{\beta}_{LS}-\beta)^2
+
p_\gamma(\beta)
\label{q_bias}
\right\}
,
\end{eqnarray}
%
%
%
%
%
%
where $\hat{\beta}_{LS}$ is the least-squares estimate in univariate case.
Figure \ref{Est_bias} shows $\hat{\beta}$ of lasso, smoothly-clipped absolute deviation (SCAD; Fan and Li, 2001), and lasso-type NEG modeling based on $(\ref{q_bias})$.
The lasso has a large bias from $\hat{\beta}_{LS}$. 
SCAD has less biased for large $|\hat{\beta}_{LS}|$.
The lasso-type modeling via the NEG distribution has a similar form to that of SCAD, but the change is continuous in $\hat{\beta}_{LS}$.

\vspace{-10mm}

%
%
%
%
%
%
%
%
%
%
%
%
%
%

%
%
%
%
%
%
\section{Bayesian fused lasso modeling via NEG prior}
%
%
%
%
%
%
\subsection{Bayesian fused lasso via NEG prior}
%
%
%
%
%
%

In this section, we propose a Bayesian sparse modeling having higher versatility than the fused lasso.
The Bayesian fused lasso assumes two independent Laplace distributions as the prior distributions for the regression coefficients $\vector{\beta}$ and their successive differences. 
By replacing the Laplace distribution for the differences with the NEG distribution, we propose the prior distribution
\begin{eqnarray}
\pi(\vector{\beta}|\sigma^2)
=
(\sigma^2)^{-(2p-1)/2}
\prod_{j=1}^p \mbox{Laplace}\left( \frac{\beta_j}{\sqrt{\sigma^2}} \Big| \lambda_1 \right)
\prod_{j=2}^p \mbox{NEG}\left( \frac{\beta_j-\beta_{j-1}}{\sqrt{\sigma^2}} \Big| \lambda_2, \gamma_2 \right)
.
\label{NEGflasso}
\end{eqnarray}
In using the NEG distribution, compared to the Laplace distribution, 
the closer the difference between two regression coefficients is, 
the stronger the penalty becomes. 
Consequently, by adding the NEG penalty for the differences for regression coefficients, the truly identical regression coefficients tend to be estimated as identical, while the truly different regression coefficients tends to be estimated as different. 

\begin{figure}[htbp]
  \centering
 \begin{minipage}{0.49\columnwidth}
  \centering
   \includegraphics[bb=0 0 734 732 ,width=80mm,clip]{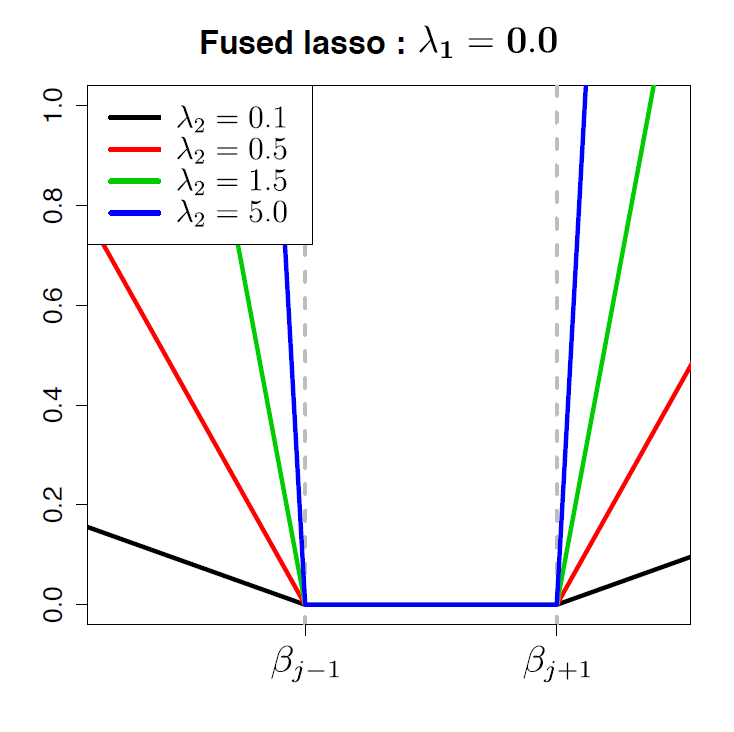}
 \end{minipage}
 \begin{minipage}{0.49\columnwidth}
  \centering
   \includegraphics[bb=0 0 734 732 ,width=80mm,clip]{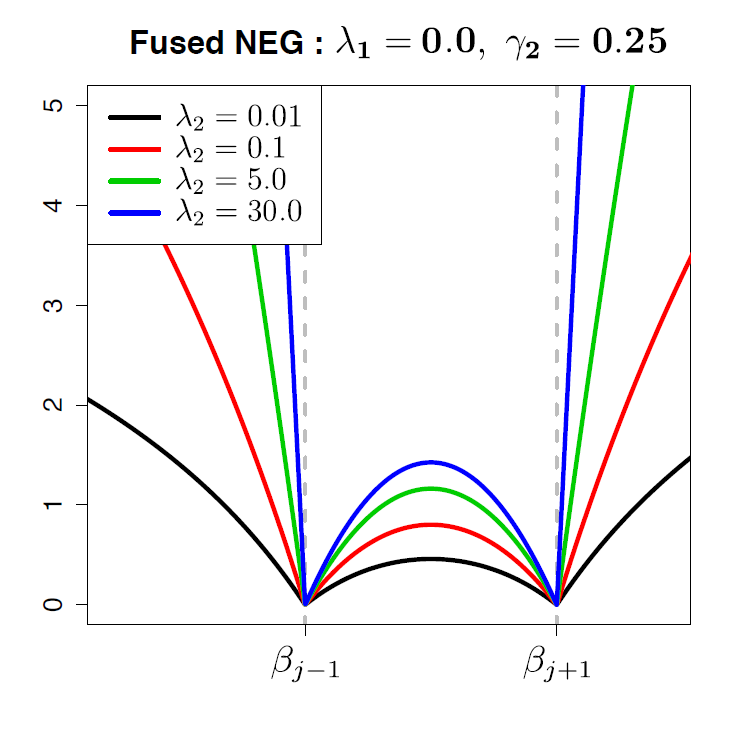}
 \end{minipage}
 \begin{minipage}{1\columnwidth}
  \centering
\vspace{-20mm}
   \includegraphics[bb=0 0 734 732 ,width=80mm,clip]{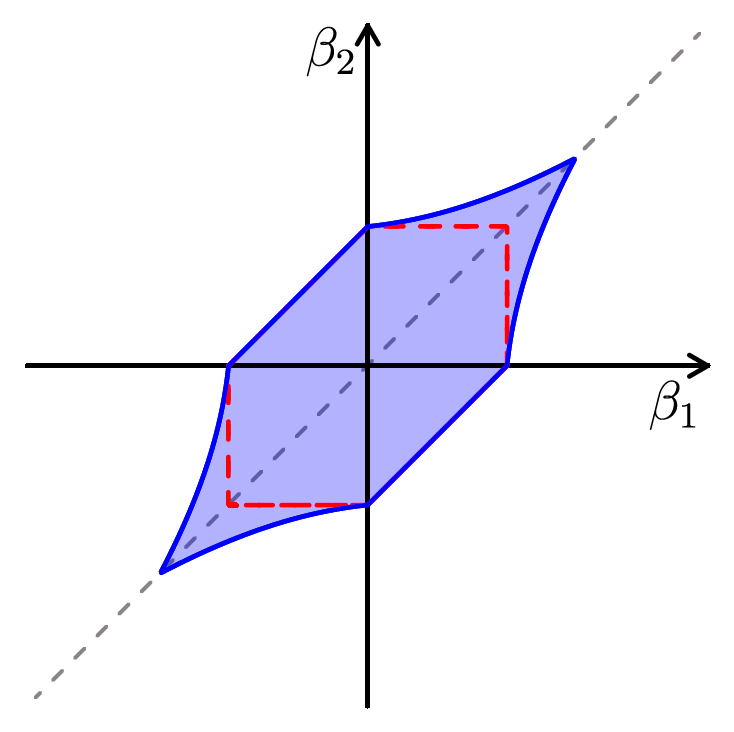}
 \end{minipage}
  \caption{
Upper left panel: The function ($\ref{fused_penalty_f}$),
$p_{\lambda_2}(\beta_j)=
\lambda_2\Big(
|\beta_j-\beta_{j-1}|+|\beta_{j+1}-\beta_j|
\Big)$, where $\beta_{j-1}$ and $\beta_{j+1}$ are fixed. 
Upper right panel: The function ($\ref{NEGfused_penalty_f}$),
$p_{\lambda_2,\gamma_2}(\beta_j)=
\log\mbox{NEG}(\beta_j-\beta_{j-1}|\lambda_2,\gamma_2)+
\log\mbox{NEG}(\beta_{j+1}-\beta_{j}|\lambda_2,\gamma_2)+
C$, where $\beta_{j-1}$ and $\beta_{j+1}$ are fixed. 
Lower panel: A constraint region of fused lasso via NEG penalty (shaded region). The red dotted line indicates fused lasso.
}
  \label{FlassoNEG_penal}
\end{figure}

The upper right panel of Figure $\ref{FlassoNEG_penal}$ shows the penalty function
\begin{eqnarray}
p_{\lambda_2,\gamma_2}(\beta_j)=
\log\mbox{NEG}(\beta_j-\beta_{j-1}|\lambda_2,\gamma_2)+
\log\mbox{NEG}(\beta_{j+1}-\beta_{j}|\lambda_2,\gamma_2)+
C
,
\label{NEGfused_penalty_f}
\end{eqnarray}
where $C$ is a constant such that $p_{\lambda_2,\gamma_2}(\beta_j)$ takes zero value at $\widetilde{\beta}=\argmin p_{\lambda_2,\gamma_2}(\beta_j)$.
When $\widetilde{\beta}$ satisfies an inequality $\beta_{j-1} \leq \widetilde{\beta} \leq \beta_{j+1}$, 
the fused lasso penalty $p_{\lambda_2}(\widetilde{\beta})$ always takes the minimum value, 
but the penalty of the proposed method does not always.
The resulting estimator based on prior ($\ref{NEGflasso}$) tends to be identical to either $\beta_{j-1}$ or $\beta_{j+1}$, and more contrasted result is obtained than the fused lasso penalty. 
This shows that the prior ($\ref{NEGflasso}$) is more flexible than that of the Bayesian fused lasso.

\vspace{-17.5mm}
%
%
%

A full-conditional distribution is obtained for each of the prior distributions, enabling Bayesian estimation by  Gibbs sampling.
The prior ($\ref{NEGflasso}$) can be expressed as a hierarchical representation
\begin{eqnarray*}
\pi(\vector{\beta}|\sigma^2)
&=&
(\sigma^2)^{-(2p-1)/2}
\prod_{j=1}^p \mbox{Laplace}\left( \frac{\beta_j}{\sqrt{\sigma^2}} \Big|\lambda_1 \right)
\prod_{j=2}^p \mbox{NEG}\left( \frac{\beta_j-\beta_{j-1}}{\sqrt{\sigma^2}}\Big|\lambda_2, \gamma_2 \right)
\\
&=& 
\int\dots\int
\prod_{j=1}^p 
\frac{1}{\sqrt{2\pi \sigma^2 \tau_j^2}} 
\exp \left( -\frac{\beta_j^2}{2\sigma^2\tau_j^2} \right) 
\prod_{j=1}^p 
\frac{\lambda_1^2}{2} \exp \left( -\frac{\lambda_1^2\tau_j^2}{2} \right)
\\
&& \qquad\quad\times
\prod_{j=2}^p 
\frac{1}{\sqrt{2\pi\sigma^2\widetilde{\tau}_j^2}}
\exp\left\{-\frac{(\beta_j-\beta_{j-1})^2}{2\sigma^2\widetilde{\tau}_j^2}\right\}
\prod_{j=2}^p 
\psi_j \exp\left(-\psi_j\widetilde{\tau}_j^2\right)
\\
&& \qquad\quad\times
\prod_{j=2}^p 
\frac{(\gamma_2^2)^{\lambda_2}}{\Gamma(\lambda_2)}\psi_j^{\lambda_2-1}\exp(-\gamma_2^2\psi_j)
\ 
\prod_{j=1}^p d\tau_j^2 \prod_{j=2}^p d\widetilde{\tau}_j^2 \prod_{j=2}^p d\psi_j
.
\end{eqnarray*}
Therefore, the priors on $\vector{\beta},\tau_1^2,\tau_2^2,\dots,\tau_p^2,\widetilde{\tau}_2^2,\widetilde{\tau}_3^2,\dots,\widetilde{\tau}_p^2,\psi_2,\psi_3,\dots,\psi_p$ are
\begin{eqnarray*}
\vector{\beta}|\sigma^2,\tau_1^2,\tau_2^2,\dots,\tau_p^2,\widetilde{\tau}_2^2,\widetilde{\tau}_3^2,\dots,\widetilde{\tau}_p^2
&\sim&
\mbox{N}_p(\vector{0}_p,\sigma^2\Sigma_{\vector{\beta}})
,
\\
\tau_j^2
&\sim&
\mbox{EXP}(\lambda_1^2/2)
,
\\
\widetilde{\tau}_j^2|\psi_j
&\sim&
\mbox{EXP}(\psi_j)
,
\\
\psi_j
&\sim&
\mbox{Ga}(\lambda_2,\gamma_2^2)
,
\end{eqnarray*}
where $\Sigma_{\vector{\beta}}$ is given by the formula ($\ref{Sigmabeta}$).
Hence the full-conditional distributions of parameters are given by
\begin{eqnarray}
\vector{\beta}|\vector{y},X,\sigma^2,\tau_1^2,\dots,\tau_p^2,\widetilde{\tau}^2_2,\dots,\widetilde{\tau}^2_p,\psi_2,\dots,\psi_p
&\sim&
\mbox{N}_p\left( A^{-1}X^T\vector{y},\ \sigma^2A^{-1} \right)
,
\nonumber
\\
&&
A=X^TX+\Sigma^{-1}_{\vector{\beta}}
,
\nonumber
\\
\sigma^2|\vector{y},X,\vector{\beta},\tau_1^2,\dots,\tau_p^2,\widetilde{\tau}^2_2,\dots,\widetilde{\tau}^2_p,\psi_2,\dots,\psi_p
&\sim&
\mbox{IG}\left( \nu_1/2,\ \eta_1/2 \right)
,
\nonumber
\\
&&
\nu_1 = n+2p-1+\nu_0
,
\nonumber
\\
&&
\eta_1 =
(\vector{y}-X\vector{\beta})^T(\vector{y}-X\vector{\beta})+\vector{\beta}^T\Sigma^{-1}_{\vector{\beta}}\vector{\beta}+\eta_0
,
\nonumber
\\
\frac{1}{\tau_j^2}|\beta_j,\sigma^2,\lambda_1
&\sim&
\mbox{IGauss}
\left( \sqrt{\frac{\lambda_1^2\sigma^2}{\beta_j^2}},\lambda_1^2 \right)
,
\nonumber
\\
\frac{1}{\widetilde{\tau}_j^2}|\beta_j,\beta_{j-1},\sigma^2,\psi_j
&\sim&
\mbox{IGauss}\left( \sqrt{\frac{2\sigma^2\psi_j}{(\beta_j-\beta_{j-1})^2}},\ 2\psi_j \right)
,
\nonumber
\\
\psi_j|\widetilde{\tau}_j^2,\lambda_2,\gamma_2
&\sim&
\mbox{Ga}\left(\lambda_2+1,\widetilde{\tau}_j^2+\gamma^2_2\right)
.
\label{fullcond_NEGflasso}
\end{eqnarray}
%
%
%
\subsection{Bayesian generalized fused lasso via NEG prior}
%
%
%

The generalized fused lasso is given by the optimization problem  
\begin{eqnarray}
\max_{\beta_1, \ldots,\beta_p}  \left\{
-\sum_{i=1}^n (y_i-\beta_i)^2
-\lambda_1\sum_{j=1}^p|\beta_j|
-\lambda_2\sum_{(k,l) \in E}|\beta_k-\beta_{l}|
\right\}
.
\end{eqnarray}
Various problems are included under this framework by changing the set $E$.
In this section, we consider using the NEG distribution for the generalized fused lasso.

%
%
%
%
%
%
%
%
%
%
%
%
%
%
\subsubsection{2d fused lasso}
%
%
%
The 2d fused lasso is a useful application of the generalized fused lasso. 
The purpose of this method is the denoising of image data.
The gray scale of $p_1\times p_2$ pixel in the image data corresponds to each $y_{i,j}$ ($i=1,\dots,p_1,\ j=1,\dots,p_2$) as shown in Figure $\ref{image_result}$.
We consider the following optimization problem:
\begin{eqnarray}
&&
\max_{\beta_{1,1}, \ldots,\beta_{p_1,p_2}} \Bigg\{
-\sum_{i=1}^{p_1}\sum_{j=1}^{p_2} (y_{i,j}-\beta_{i,j})^2
-\lambda_1\sum_{i=1}^{p_1}\sum_{j=1}^{p_2}|\beta_{i,j}|
\nonumber\\ 
&&\qquad\qquad
-\lambda_2\sum_{i=1}^{p_1}\sum_{j=2}^{p_2}|\beta_{i,j}-\beta_{i,j-1}|
-\lambda_2\sum_{i=2}^{p_1}\sum_{j=1}^{p_2}|\beta_{i,j}-\beta_{i-1,j}|
\Bigg\}
.
\end{eqnarray}
The estimated value of parameter $\beta_{ij}$ corresponds to the denoised image.

Next, we formulate the 2d fused lasso in a Bayesian framework.
For the following discussions, we use the notations
\begin{eqnarray*}
\vector{y}
&=&
(y_{1,1},\dots,y_{1,p_2},y_{2,1},\dots,y_{2,p_2},\dots,y_{p_1,1},\dots,y_{p_1,p_2})^T\\
&=&
(y_1,y_2,\cdots,y_p)^T
,
\\
\vector{\beta}
&=&
(\beta_{1,1},\dots,\beta_{1,p_2},\beta_{2,1},\dots,\beta_{2,p_2},\dots,\beta_{p_1,1},\dots,\beta_{p_1,p_2})^T\\
&=&
(\beta_1,\beta_2,\cdots,\beta_p)^T
,
\end{eqnarray*}
where $p=p_1\times p_2$.
The likelihood function and prior distribution on $\vector{\beta}$ are, respectively,
\begin{eqnarray}
f(\vector{y}|\vector{\beta},\sigma^2)
&=&
\mbox{N}_p(\vector{\beta},\sigma^2I_p)
,
\\
\pi(\vector{\beta}|\sigma^2)
&\propto&
(\sigma^2)^{-(3p-p_1-p_2)/2}
\prod_{j=1}^p \frac{\lambda_1}{2}\exp\left(
-\frac{\lambda_1}{\sigma}|\beta_j|
\right)
\nonumber \\
&&\qquad
\times
\prod_{j \in \Omega_1}
\mbox{NEG}(\beta_j-\beta_{j-1}|\lambda_2,\gamma_2)
\prod_{j \in \Omega_2}
\mbox{NEG}(\beta_{j}-\beta_{j-p_2}|\lambda_2,\gamma_2)
\label{2dNEG_piror}
,
\end{eqnarray}
where $\Omega_1=\{1,2,\dots,p\}\backslash \{1,p_2+1,\dots,(p_1-1)p_2+1\},\ \Omega_2=\{p_2+1,p_2+2,\dots,p\}$.
The prior ($\ref{2dNEG_piror}$) can be expressed as a hierarchical representation
\begin{eqnarray*}
\pi(\vector{\beta}|\sigma^2)
&=& 
\int\dots\int
\prod_{j=1}^p 
\frac{1}{\sqrt{2\pi \sigma^2 \tau_j^2}} 
\exp \left( -\frac{\beta_j^2}{2\sigma^2\tau_j^2} \right) 
\prod_{j=1}^p 
\frac{\lambda_1^2}{2} \exp \left( -\frac{\lambda_1^2\tau_j^2}{2} \right)
\\
&& \qquad\quad\times
\prod_{j \in \Omega_1}
\frac{1}{\sqrt{2\pi\sigma^2\widetilde{\tau}_{j-1,j}^2}}\exp\left\{-\frac{(\beta_j-\beta_{j-1})^2}{2\sigma^2\widetilde{\tau}_{j-1,j}^2}\right\}
\prod_{j \in \Omega_1} 
\psi_{j-1,j} \exp\left(-\psi_{j-1,j}\widetilde{\tau}_{j-1,j}^2\right)
\\
&&\qquad\quad\times
\prod_{j \in \Omega_1}
\frac{(\gamma_2^2)^{\lambda_2}}{\Gamma(\lambda_2)}\psi_{j-1,j}^{\lambda_2-1}\exp(-\gamma_2^2\psi_{j-1,j})
\\
&& \qquad\quad\times
\prod_{j \in \Omega_2}
\frac{1}{\sqrt{2\pi\sigma^2\widetilde{\tau}_{j-p_2,j}^2}}\exp\left\{-\frac{(\beta_j-\beta_{j-p_2})^2}{2\sigma^2\widetilde{\tau}_{j-p_2,j}^2}\right\}
\prod_{j \in \Omega_2} 
\psi_{j-p_2,j} \exp\left(-\psi_{j-p_2,j}\widetilde{\tau}_{j-p_2,j}^2\right)
\\
&&\qquad\quad\times
\prod_{j \in \Omega_2}
\frac{(\gamma_2^2)^{\lambda_2}}{\Gamma(\lambda_2)}\psi_{j-p_2,j}^{\lambda_2-1}
\exp\left(-\gamma_2^2\psi_{j-p_2,j}\right)
\\ 
&& \qquad\quad\times
\prod_{j=1}^p d\tau_j^2 
\prod_{j \in \Omega_1} d\widetilde{\tau}_{j-1,j}^2 \prod_{j \in \Omega_1} d\psi_{j-1,j}
\prod_{j \in \Omega_2} d\widetilde{\tau}_{j-p_2,j}^2 \prod_{j \in \Omega_2} d\psi_{j-p_2,j}
.
\end{eqnarray*}
The full-conditional distribution is then obtained by replacing $\Sigma_{\vector{\beta}}^{-1}$ by the following expression in the fused lasso-type Bayesian modeling via the NEG distribution in Equation ($\ref{fullcond_NEGflasso}$):
\begin{eqnarray*}
(\Sigma_{\vector{\beta}}^{-1})_{(i,j)}
=
\begin{cases}
\displaystyle
\frac{1}{\tau^2_i}+
\frac{1}{\widetilde{\tau}^2_{i-1,j}}+\frac{1}{\widetilde{\tau}^2_{i-p_2,j}}
+
\frac{1}{\widetilde{\tau}^2_{i,j+1}}+\frac{1}{\widetilde{\tau}^2_{i,j+p_2}}
& 
(i=j)
\\
\displaystyle
-\frac{1}{\widetilde{\tau}^2_{i,j}}
&
(j\in \{i+1,\ i+p_2,i-1,i-p_2\})
\\
\displaystyle
0 & (\mbox{otherwise})
\end{cases}
\end{eqnarray*}
where $(\Sigma_{\vector{\beta}}^{-1})_{(i,j)}$ is the $(i,j)$-element of $\Sigma_{\vector{\beta}}^{-1}$ and $1/\widetilde{\tau}^2_{i,j}=1/\widetilde{\tau}^2_{j,i}$,
 $1/\widetilde{\tau}^2_{j'-1,j'}=0$ ( $ j' \in \{1,\dots,p \}\setminus \Omega_1 $),
$1/\widetilde{\tau}^2_{j'-p_2,j'}=0$ ($ j' \in \{ 1,\dots,p \} \setminus \Omega_2 $).
%
%
%

%
%
%
%
%
%
%
%
%
\subsubsection{HORSES}
%
%
%
In the fused lasso, the predictors must be in some sense ordered. 
On the other hand, HORSES does not have such a requirement.
In the HORSES, all pairwise differences of two regression coefficients are used as a penalty.
The regularization method maximizes the objective function
\begin{eqnarray}
\log f(\vector{y}|\vector{\beta},\sigma^2)
-\lambda_1\sum_{j=1}^p |\beta_j|
-\lambda_2\sum_{j>k} |\beta_j-\beta_k|
.
\end{eqnarray}

Next, we formulate HORSES in a Bayesian framework. The prior on $\vector{\beta}$ is assumed as
\[
\pi(\vector{\beta}|\sigma^2)
=
(\sigma^2)^{-(p+p(p-1)/2)/2}
\prod_{j=1}^p \mbox{Laplace}\left( \frac{\beta_j}{\sqrt{\sigma^2}} \Big| \lambda_1 \right)
\prod_{j>k} \mbox{NEG}\left( \frac{\beta_j-\beta_k}{\sqrt{\sigma^2}} \Big| \lambda_2, \gamma_2 \right)
.
\]
The full-conditional distribution is obtained by replacing the $p\times p$ matrix $\Sigma_{\vector{\beta}}$ in the fused lasso-type Bayesian modeling via an NEG distribution in ($\ref{fullcond_NEGflasso}$) by
\begin{eqnarray*}
(\Sigma_{\vector{\beta}}^{-1})_{(i,j)}
=
\begin{cases}
\displaystyle
\frac{1}{\tau^2_i}+\sum_{j' \neq i} \frac{1}{\widetilde{\tau}^2_{i,j'}}
& 
(i=j)
\\
\displaystyle
-\frac{1}{\widetilde{\tau}^2_{i,j}}
&
(\mbox{otherwise})
\end{cases}
,
\end{eqnarray*}
where $(\Sigma_{\vector{\beta}}^{-1})_{(i,j)}$ is the $(i,j)$-element of $\Sigma_{\vector{\beta}}^{-1}$.

%
%
%
%
%
%
\subsection{Computational algorithm for exact sparse solution}
%
%
%

Since a posterior mode is estimated by random numbers, the Gibbs sampling does not produce exact zero estimates of the coefficients. 
The fused lasso has two purposes: sparse estimation of both the coefficients and differences between adjacent regression coefficients. 
To achieve these two purposes,
we propose the Sparse Fused Algorithm (SFA), 
which allows both regression coefficients and differences of regression coefficients to be exactly zero.
The details of the algorithm are given in Table \ref{SFAalgorithm}.
By modifying this algorithm slightly, we can also construct an algorithm for the generalized fused lasso. 

\begin{table}[htbp]
\caption{Sparse Fused algorithm (SFA)}
\begin{center}
 \newcounter{tabenum}\setcounter{tabenum}{0}
 \newcommand{\nextnum}{\addtocounter{tabenum}{1}\thetabenum. }
\begin{tabular}{l|l}
\hline
\multicolumn{2}{l}{
\vspace{-10pt}
}\\
\shortstack[l]{
\nextnum 
Let 
$\hat{\vector{\beta}}=(\hat{\beta}_1,\dots,\hat{\beta}_p)^T$ \\
\hspace{9pt}
be a vector of estimates obtained\\
\hspace{9pt}
from Gibbs sampling.\\
\hspace{9pt}
$\vector{I}=(I_1,I_2,\dots,I_p) \leftarrow (1,2,\dots,p)$ \\
\nextnum 
$\widetilde{\vector{\beta}}=(\widetilde{\beta}_1,\widetilde{\beta}_2,\dots,\widetilde{\beta}_p)^T \leftarrow \hat{\vector{\beta}}$ \\
\hspace{9pt}
$\widetilde{\vector{\beta}}^{(f)}=(\widetilde{\beta}_1^{(f)},\widetilde{\beta}_2^{(f)},\dots,\widetilde{\beta}_p^{(f)})^T \leftarrow \hat{\vector{\beta}}$ \\
\hspace{9pt}
$\widetilde{\vector{\beta}}^{(b)}=(\widetilde{\beta}_1^{(b)},\widetilde{\beta}_2^{(b)},\dots,\widetilde{\beta}_p^{(b)})^T \leftarrow \hat{\vector{\beta}}$ \\
\hspace{9pt}
$\widetilde{\vector{\beta}}^{(z)}=(\widetilde{\beta}_1^{(z)},\widetilde{\beta}_2^{(z)},\dots,\widetilde{\beta}_p^{(z)})^T \leftarrow \hat{\vector{\beta}}$ \\
\nextnum FOR $j=1,\dots,p$ \\
\hspace{19pt}
FOR $k=1,\dots,p$ \\
\hspace{29pt}
IF $I_k=j$ THEN \\
\hspace{39pt}
SET $\widetilde{\beta}_k^{(f)} \leftarrow \hat{\beta}_{j-1}$ \\
\hspace{39pt}
SET $\widetilde{\beta}_k^{(b)} \leftarrow \hat{\beta}_{j+1}$ \\
\hspace{39pt}
SET $\widetilde{\beta}_k^{(z)} \leftarrow 0$ \\
\hspace{29pt}
END IF \\
\hspace{19pt}
END FOR \\
\hspace{2pt}
3.1
\hspace{-2pt}
$G = g(\tilde{\vector{\beta}},\ \hat{\vector{\xi}},\ \vector{y})$ \\
\hspace{19pt}
$G_{(f)} = g(\widetilde{\vector{\beta}}^{(f)},\ \hat{\vector{\xi}},\ \vector{y})$ \\
\hspace{19pt}
$G_{(b)} = g(\widetilde{\vector{\beta}}^{(b)},\ \hat{\vector{\xi}},\ \vector{y})$ \\
\hspace{19pt}
$G_{(z)} = g(\widetilde{\vector{\beta}}^{(z)},\ \hat{\vector{\xi}},\ \vector{y})$ \\
\hspace{19pt}
${\mathcal G} = \left\{G,\ G_{(f)},\ G_{(b)},\ G_{(z)}\right\}$ 
\hspace{30pt}
}
\vspace{5pt} &
\shortstack[l]{
\hspace{2pt}
3.2
\hspace{-2pt}
FOR $k=1,\dots,p$\\
\hspace{29pt}
IF $I_k=j$ THEN \\
\hspace{39pt}
CASE $\max \left\{ {\mathcal G} \right\}$ OF\\
\hspace{49pt}
$G$：$\hat{\beta}_k \leftarrow \widetilde{\beta}_j$ \\
\hspace{49pt}
$G_{(f)}$：$\hat{\beta}_k \leftarrow \widetilde{\beta}_{j-1}$ \\
\hspace{79pt}
$I_k \leftarrow j-1$ \\
\hspace{49pt}
$G_{(b)}$：$\hat{\beta}_k \leftarrow \widetilde{\beta}_{j+1}$ \\
\hspace{77pt}
$I_k \leftarrow j+1$ \\
\hspace{49pt}
$G_{(z)}$：$\hat{\beta}_k \leftarrow 0$ \\
\hspace{78pt}
$I_k \leftarrow 0$ \\
\hspace{39pt}
END CASE \\
\hspace{29pt}
END IF \\
\hspace{19pt}
END FOR \\
\hspace{9pt}
END FOR \\ 
\nextnum 
Repeat Steps 2 and 3 until convergence \\
\hspace{9pt}
and sparsified estimates are stored in $\hat{\vector{\beta}}$.
\vspace{110pt}
}\\
\hline
\multicolumn{2}{l}{
\shortstack[l]{
\vspace{5pt}\\
Here, $g(\vector{\beta},\ \vector{\xi},\ \vector{y})=\log f(\vector{y}|\vector{\beta},\ \vector{\xi})+\log\pi(\vector{\beta},\ \vector{\xi})$,\ 
$f(\vector{y}|\vector{\beta},\ \vector{\xi})$ is a likelihood function，\\
$\pi(\vector{\beta},\ \vector{\xi})$ is a prior on $(\vector{\beta},\ \vector{\xi})$，
$\hat{\vector{\xi}}$ is an estimate of parameter vector $\vector{\xi}$ other than $\vector{\beta}$ such as $\sigma^2$.
}
}\\
\end{tabular}
\label{SFAalgorithm}
\end{center}
\end{table}
%
%
%
%
%
%

%
%
%
\subsection{Model selection}
%
%
%

Chen and Chen (2008) proposed an extended Bayesian information criterion (EBIC) to overcome the difficulties in model selection for small sample and high-dimensional data frequently encountered in genomic studies and image analysis.

The basic idea of EBIC is as follows. 
Suppose that the likelihood function is $L_n(\vector{\theta})=f(y|\vector{\theta})=\prod_{i=1}^nf(y_i|x_i,\vector{\theta})$, where $\vector{\theta} \in \Theta \subset R^p$. 
A model $M$ is a subset of $\{1,\dots,p\}$. 
It indicates indexes of variables included in the model. 
For $M$ included in the model space $\mathcal{M}$, the posterior of $M$ is given by
\[
p(M|Y)=
\frac{m(Y|M)p(M)}{{\sum_{M\in\mathcal{M}}m(Y|M)p(M)}}
,
\]
where 
$m(Y|M)$ is the marginal likelihood and $p(M)$ is the prior of $M$.
The marginal likelihood is
\[
m(Y|M)=
\int
f\big\{Y|\vector{\vector{\theta}}(M)\big\}\pi\big\{\vector{\vector{\theta}}(M)\big\}
d\vector{\theta}(M)
,
\]
where $\pi\{\vector{\theta}(M)\}$ is the prior of $\vector{\theta}(M)$ being the parameter $\vector{\theta}$ of the model $M$.
By the Laplace approximation for integrals in the above quantity, we derive
\[
-2 \log m(Y|M)
=
-2 \log L_n \{ \hat{\vector{\vector{\theta}}}(M) \}
+\nu(M)\log n
-2p(M)
,
\]
where 
$\hat{\vector{\theta}}(M)$ is the maximum likelihood estimator of $\vector{\theta}(M)$, $\nu(M)$ is the degrees of freedom of $M$. 
In addition,  terms of smaller order than $O(1)$ with respect to the sample size $n$ are ignored.
The BIC (Schwarz, 1978) approximates the posterior probability of a model by assuming that the prior is uniform over all models, and is of the form
\[
\mbox{BIC}(M)=
-2\log L_n\big\{\hat{\vector{\theta}}(M)\big\}
+\nu(M)\log n
.
\]

On the other hand, the EBIC considers the prior probability on a model $M$ which takes the number of candidate models into consideration, rather assuming a uniform prior.
Suppose that a model space $\mathcal{M}$ is partitioned into $\coprod_j \mathcal{M}_j$.
The EBIC is then given by, for $M\in\mathcal{M}_j$,
\[
\mbox{EBIC}(M)=
-2\log L_n\big\{\hat{\vector{\theta}}(M)\big\}
+\nu(M)\log n+2\gamma\log\tau(\mathcal{M}_j)
,
\]
where 
$\gamma \ (0<\gamma<1)$ is the parameter 
and
$\tau(\mathcal{M}_j)$ is a quantity which characterizes $\mathcal{M}_j$. 
Chen and Chen (2008) used $\tau(\mathcal{M}_j)= \binom{p}{j} = p!/\{ (p-j)! j! \}$ for lasso-type modeling. 
Tibshirani {\it et al}. (2005) proposed, as the degrees of freedom, 
\[
\mbox{df}(\hat{\vector{\beta}})
=\#
\left\{
\mbox{nonzero coefficient blocks in }\hat{\vector{\beta}}
\right\}
.
\]
It can be rewritten as
\[
\mbox{df}(\hat{\vector{\beta}})
=
p-
\#\left\{\hat{\beta}_j=0\right\}
-
\#\left\{\hat{\beta}_j=\hat{\beta}_{j-1};\hat{\beta}_j,\hat{\beta}_{j-1}\neq0\right\}
.
\]

In this paper, we use $\mbox{df}(\hat{\vector{y}})$ to indicate the degrees of freedom of components $\nu(M)$ in the EBIC and 
$\tau(\mathcal{M}_j)=\binom{p_g}{\mbox{df}(\hat{\vector{\beta}})} = p_g ! / [  \{ p_g - \mbox{df}(\hat{\vector{\beta}}) \} ! \mbox{df}(\hat{\vector{\beta}}) ! ] $, 
where $p_g$ is the number of coefficient blocks in $\hat{\vector{\beta}}$ including zero coefficients. 
We also use $\gamma=1-\log n/(2\log p)$ as recommended by Chen and Chen (2008).

%
%
%
\section{Numerical studies}
%
%
%

%
%
%
\subsection{Monte Carlo simulation}
%
%
%

We simulated data from the model with $n$ observations and $p$ predictors:
\[
\vector{y}=X \vector{\beta}^*+\vector{\epsilon},
\]
where $\vector{\beta}^*$ is the $p$-dimensional true coefficient vector, $\vector{\epsilon}$ is an error vector distributed as ${\rm N} _n(\vector{0}_n,\sigma^2I_n)$. 
In addition, $\vector{x}_i \ (i=1,2,\ldots,n) $ was generated from a multivariate normal distribution with mean vector $\vector{0}_p$ and variance-covariance matrix $\Sigma$. 
We simulated $200$ datasets with $n$ observations. 
We considered the following three cases.
\begin{itemize}
\item
Case 1: 
$n=50,p=20$,
$\vector{\beta}^*=(\vector{0.0}_5^T, \vector{2.0}_5^T, \vector{0.0}_5^T, \vector{2.0}_5^T)^T$, $\sigma=0.75$, $\Sigma_{ii}=1$, and $\Sigma_{ij}=0.5 (i\neq j)$, where $\Sigma_{ij}$ is the $(i,j)$-element of $\Sigma$. 
\item
Case 2: 
$n=50,p=50$,
$\vector{\beta}^*=(\vector{0.0}_5^T, \vector{5.0}_3^T, \vector{0.0}_{15}^T, \vector{3.5}_7^T, \vector{0.0}_{10}^T, \vector{4.5}_5^T, \vector{0.0}_5^T)^T$, $\sigma=0.75$, and $\Sigma=I_p$. 
\item
Case 3: 
$n=30,p=50$,
$\vector{\beta}^*=(\vector{3.0}_5^T, -\vector{1.5}_5^T, \vector{1.0}_5^T, \vector{2.0}_5^T, \vector{0.0}_{30}^T)^T$, $\sigma=5.0$, and $\Sigma_{ij}=0.5^{|i-j|}$. 
\end{itemize}

We denote the blocks of indexes which have distinctive regression coefficients by $B_1,B_2,\ldots,B_L \subset \{1,2,\ldots,p\}$.
For example, $L=4$ in Case 1.
For each generated dataset, the estimates were obtained by using 5,000 iterations of Gibbs sampler (after 2,000 burn-in iterations).
The hyper-parameter $\lambda$ was tested for 100 values; $\lambda_i=\lambda_{{\min}}\exp\{(\log \lambda_{{\max}} -\log \lambda_{{\min}})\cdot (i/100)\}\ (i=1,\dots,100),$ where $\lambda_{{\min}}=10^{-4}$ and $\lambda_{{\max}}$ is such that all coefficient parameters are zero.

We compared the lasso and fused lasso as competitors. 
The regularization parameter in the lasso was selected by 10-fold cross-validation.
Regularization parameters in the fused lasso and the proposed method were selected by the EBIC.

The performances were evaluated in terms of two accuracies: variable selection and prediction.
For variable selection accuracy, we used three measures:
\begin{eqnarray*}
P_{\rm{Z}} &=& \frac{1}{200} \sum_{k=1}^{200} 
\frac{\#\{j:\beta_j^{(k)}=0\land \beta_j^* =0\}}{\#\{j:\beta_j^*=0\}}, \\
P_{\rm{NZ}} &=& \frac{1}{200} \sum_{k=1}^{200} 
\frac{\#\{j:\beta_j^{(k)}\neq0\land \beta_j^* \neq0\}}{\#\{j:\beta_j^*\neq0\}} ,\\
P_{\rm{B}} &=& \frac{1}{200}\sum_{k=1}^{200} 
\frac{
p-\sum_{l=1}^L
N_l^{(k)}
}{p-L},
\end{eqnarray*}
where $\hat{\vector{\beta}}^{(k)}=(\hat{\beta}_1^{(k)},\dots,\hat{\beta}_p^{(k)})^T$ is the estimate of coefficient vector for the $k$-th dataset,
and $N_l^{(k)}$ is the number of distinct regression coefficients $\{ \hat{\beta}_j^{(k)} : j \in B_l \}$.
$P_{\rm{Z}}$ indicates the accuracy of identifying truly zero coefficients.
$P_{\rm{NZ}}$ indicates the accuracy of identifying truly nonzero coefficients.
$P_{\rm{B}}$ indicates the accuracy of identifying the true coefficient blocks.
The higher the value, the more accurate variable selection is.
We assessed the accuracy of prediction using the mean squared error (MSE) and prediction squared error (PSE) as follows:
\begin{eqnarray*}
\mbox{MSE}&=&\frac{1}{200}\sum_{k=1}^{200}
(\hat{\vector{\beta}}^{(k)}-\vector{\beta}^*)^T\Sigma(\hat{\vector{\beta}}^{(k)}-\vector{\beta}^*)
,
\\
%
\mbox{PSE}&=& \frac{1}{200}\sum_{k=1}^{200}
\left(\frac{1}{n}
\|\hat{\vector{y}}^{(k)}-\widetilde{\vector{y}}^{(k)}\|^2_2
\right)
,
\end{eqnarray*}
where $\widetilde{y}^{(k)} = X^{(k)}\vector{\beta}^*+\widetilde{\vector{\epsilon}}^{(k)}$, 
with $\widetilde{\vector{\epsilon}}^{(k)}$ being an observation independent of $k$-th error vector $\vector{\epsilon}^{(k)}$.

The simulation results are summarized in Table $\ref{table1}$.
First, the lasso shows low $P_{\rm{B}}$
because it can not handle regression coefficients as blocks, and blocks of zero coefficients exist.
The fused lasso outperformed the lasso because of accounting for the block structure.
Irrespective of accuracy criteria,
the proposed method showed much better performance than those of the compared methods.
This demonstrates that the true blocks were almost identified by the proposed method, as seen in the value of $P_{\rm{B}}$ being close to $1$. 
Moreover, the fact that both MSE and PSE were low shows that our method enables proper estimates of not only the true blocks but also their true regression coefficients.
\begin{table}[htbp]
\begin{center}
\caption{The results for Monte Carlo simulations. flasso indicates fused lasso. NEG-flasso indicates our proposed fused lasso-type modeling via the NEG prior distribution.}\label{table1}
\vspace{3mm}
\begin{tabular}{cccccccc}
\hline
&\multicolumn{7}{c}{Case 1 : $n=50,\ p=20$}\\
\cline{2-8}
 &MSE & (sd) & PSE & (sd) & $P_{\rm{Z}}$ & $P_{\rm{NZ}}$ & $P_{\rm{B}}$ \\
\hline
lasso		& 0.49	&	(0.27)	&	0.83	&	(0.20)	&	0.64	&	1.00	&	0.28	\\
flasso		& 0.27	&	(0.20)	&	0.69	&	(0.15)	&	0.49	&	1.00	&	0.89	\\
NEG-flasso	& 0.03	&	(0.05)	&	0.59	&	(0.12)	&	0.96	&	1.00	&	1.00	\\
\hline
\end{tabular}
\end{center}
\vspace{-4mm}
\begin{center}
\vspace{3mm}
\begin{tabular}{cccccccc}
\hline
&\multicolumn{7}{c}{Case 2 : $n=50,\ p=50$}\\
\cline{2-8}
 &MSE & (sd) & PSE & (sd) & $P_{\rm{Z}}$ & $P_{\rm{NZ}}$ & $P_{\rm{B}}$ \\
\hline
lasso		& 1.37	& (0.83)	& 1.01	& (0.22)	& 0.61	& 1.00	& 0.40	\\
flasso		& 0.46	& (0.24)	& 0.88	& (0.20)	& 0.74	& 1.00	& 0.89	\\
NEG-flasso	& 0.04	& (0.03)	& 0.60	& (0.12)	& 1.00	& 1.00	& 1.00	\\
\hline
\end{tabular}
\end{center}
\vspace{-4mm}
\begin{center}
\vspace{3mm}
\begin{tabular}{cccccccc}
\hline
&\multicolumn{7}{c}{Case 3 : $n=30,\ p=50$}\\
\cline{2-8}
 &MSE & (sd) & PSE & (sd) & $P_{\rm{Z}}$ & $P_{\rm{NZ}}$ & $P_{\rm{B}}$ \\
\hline
lasso		& 57.83	& (14.75)		& 60.29	& (28.82)	& 0.87	& 0.47	& 0.71	\\
flasso		& 76.38	& (36.55)		& 48.56	& (12.40)	& 0.28	& 0.86	& 0.47	\\
NEG-flasso	& 10.54	& (8.92)		& 35.81	& (10.56)	& 0.49	& 0.96	& 0.94	\\
\hline
\end{tabular}
\end{center}
\end{table}
%
%
%
%
%
%
%
%
%
%
%
%
%

%
%
%
\subsection{Demonstration with artificial data for FLSA model}
\label{Illustration_through_artificial_data_for_FLSA_model}
%
%
%
We demonstrated our proposed method with artificial data generated from the FLSA model
\begin{eqnarray}
\vector{y}=\vector{\beta}^*+\vector{\epsilon},
\end{eqnarray}
where $\vector{\beta}^*$ is the $p$-dimensional true parameter and $\vector{\epsilon} \sim \mbox{N}_p(\vector{0}_p,\sigma^2I_p)$. 
We considered $\vector{\beta}^*=(-\vector{1}_5^T,\vector{0}_{20}^T,\vector{2}_5^T,\vector{0}_{40}^T,\vector{4}_{10}^T,\vector{0}_5^T,\vector{2}_5^T,\vector{0}_{10}^T)^T$ and $\sigma=0.5$. 
The hyper-parameters $(\lambda_1,\lambda_2,\gamma_2)$ were tested for $(200,200,5)$ candidate values and chosen by the EBIC. We used the fused lasso as a competitor.

Figure $\ref{res_flsa}$ gives estimates from the proposed method and the fused lasso. 
It can be seen that the proposed method estimates the true blocks more accurately than the fused lasso. 
In the fused lasso, the blocks of nonzero coefficients estimated by the fused lasso have been largely shrunken toward zero.
As a consequence, the estimated values were highly biased from the true values.
On the other hand, the proposed method could successfully estimate the true coefficients blocks. 
The proposed method gave no blocks consisting of single coefficient,
while the fused lasso had such seven blocks. 
In this illustration, our proposed method also captured the true structure better than the fused lasso. 
%
%
%
\subsection{Comparative genomic hybridization analysis for FLSA model}
%
%
%

We applied our proposed method to a real dataset;
comparative genomic hybridization (CGH) data. 
The dataset was taken from the {\tt cghFLasso} package in the software \texttt{R}. 
We randomly extracted $110$ samples from the dataset. 
We compared the proposed method to the FLSA procedure of Tibshirani and Wang (2008), which is implemented in the {\tt cghFLasso} package. 

\begin{figure}[htbp]
\vspace{-30.0mm}
  \begin{center}
   \includegraphics[bb=0 0 665 664 ,width=110mm,clip]{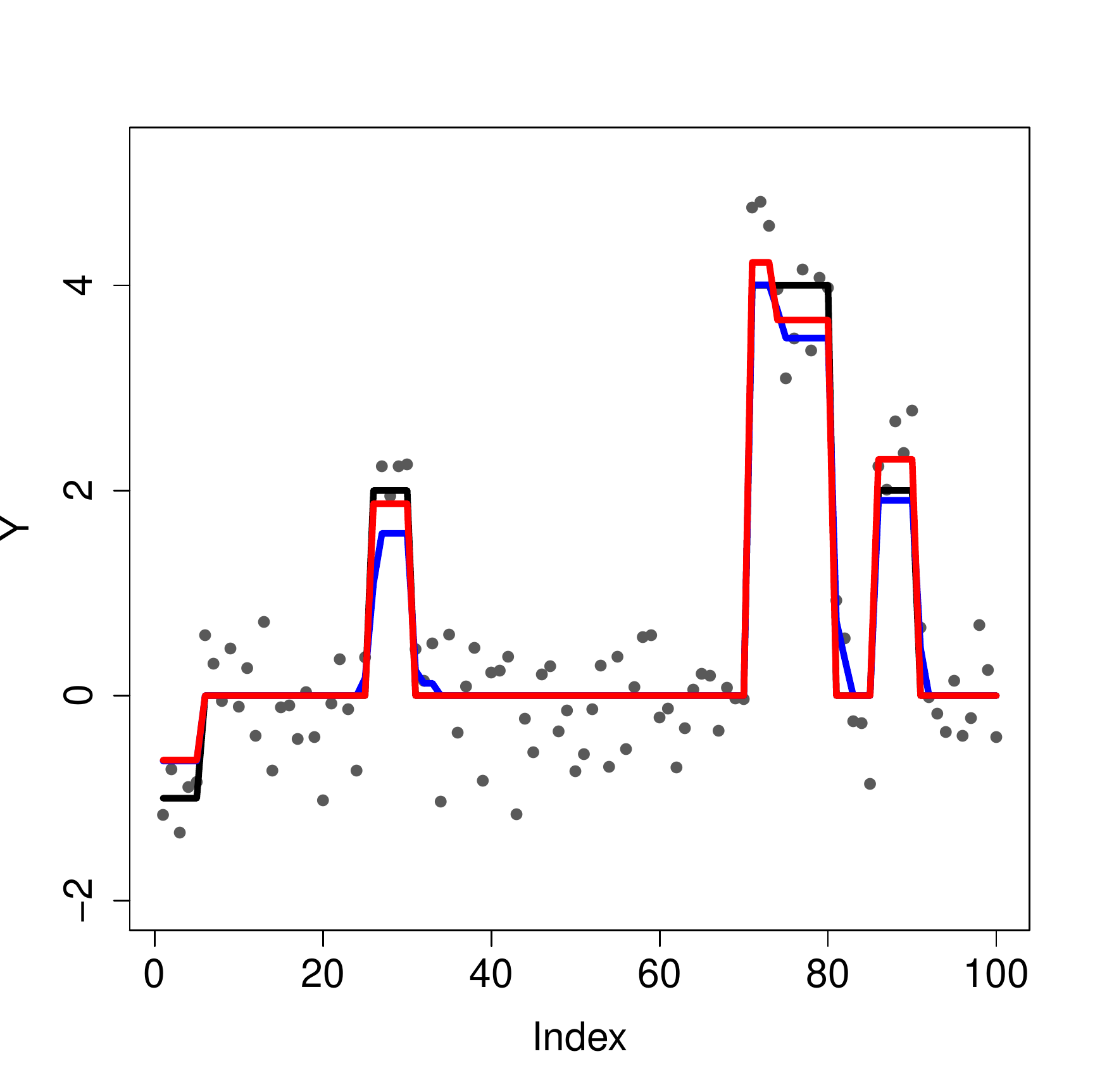}
  \end{center}
  \vspace{-5.0mm}
  \caption{The result for the simulation in Section \ref{Illustration_through_artificial_data_for_FLSA_model}. 
  Black dots indicate the simulated data, the black line is the true model, the blue line is the estimator of fused lasso, and the red line is the estimator of the proposed method.}
  \label{res_flsa}
\vspace{-30.0mm}
  \begin{center}
   \includegraphics[bb=0 0 665 664 ,width=110mm,clip]{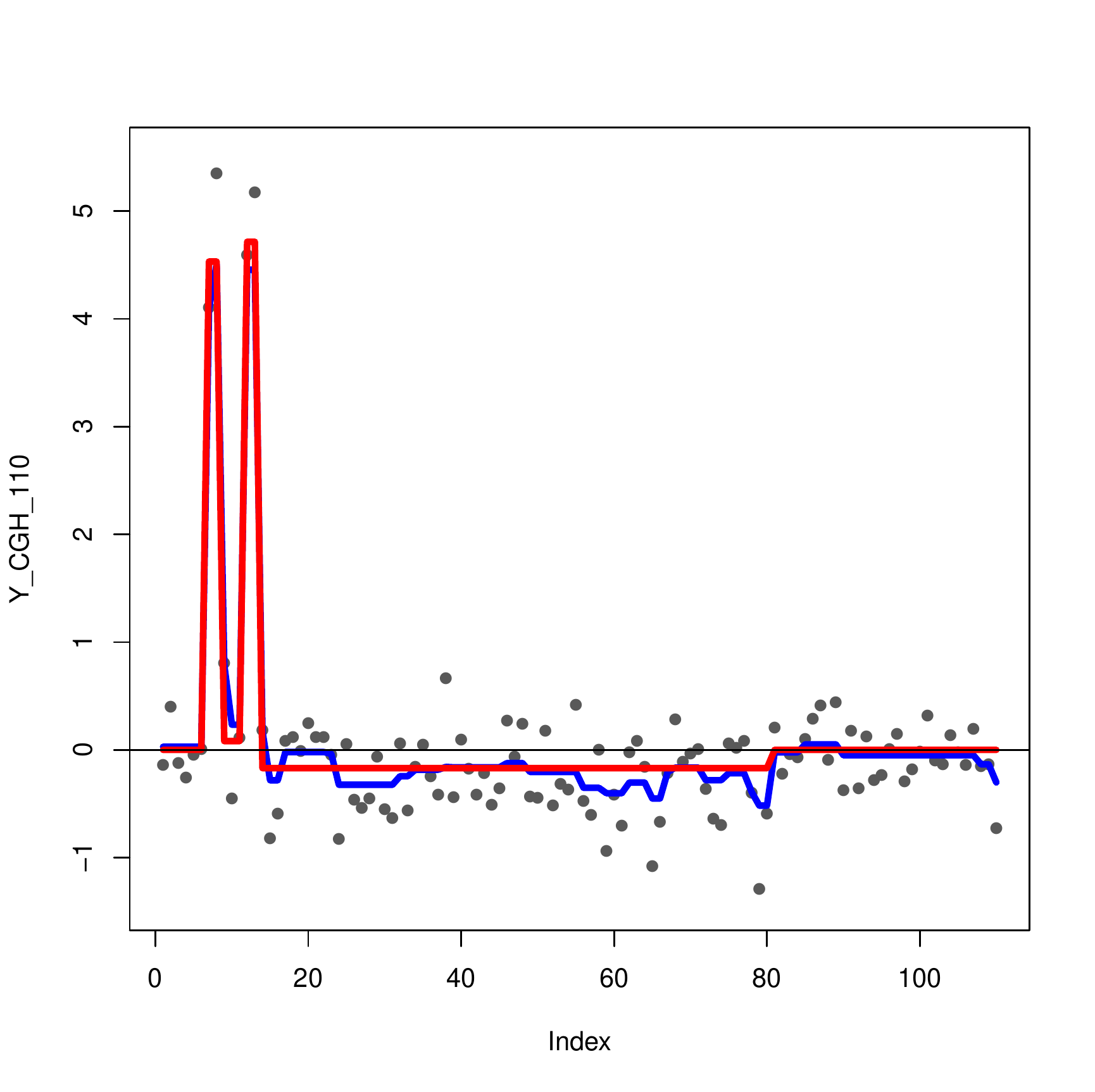}
  \end{center}
  \vspace{-5.0mm}
  \caption{The result for the comparative genome hybridization (CGH) analysis. Black dots indicate data points, the blue line is the estimator of fused lasso,  and the red line is the estimator of the proposed method.}
  \label{res_cgh}
\end{figure}

Figure $\ref{res_cgh}$ gives the result of real data analysis.
The FLSA procedure provided seemingly an over-fitted model, that is,
the estimated model existed overly close to the data. 
On the other hand, the proposed method seemed to give a more clear-cut estimate.
%
%
%
%
%
%
%
%
%
\subsection{Demonstration with artificial data for 2d fused lasso model}
%
%
%
%
%
%
%

\begin{figure}[htbp]
  \centering
 \begin{minipage}{0.49\columnwidth}
  \centering
   \includegraphics[bb=0 0 849 732 ,width=120mm,clip]{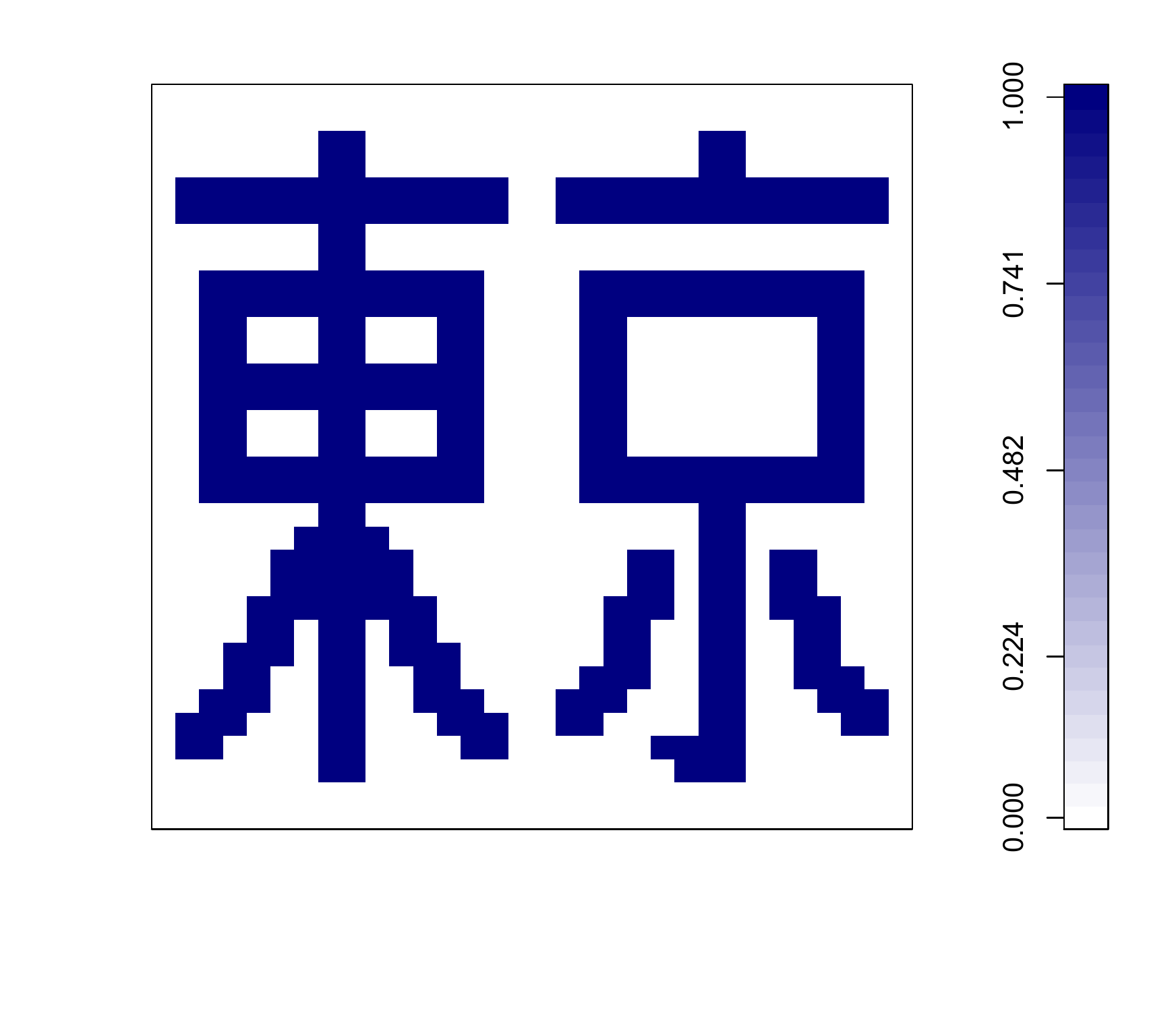}
 \end{minipage}
\vspace{-50mm}
 \begin{minipage}{0.49\columnwidth}
  \centering
   \includegraphics[bb=0 0 849 732 ,width=120mm,clip]{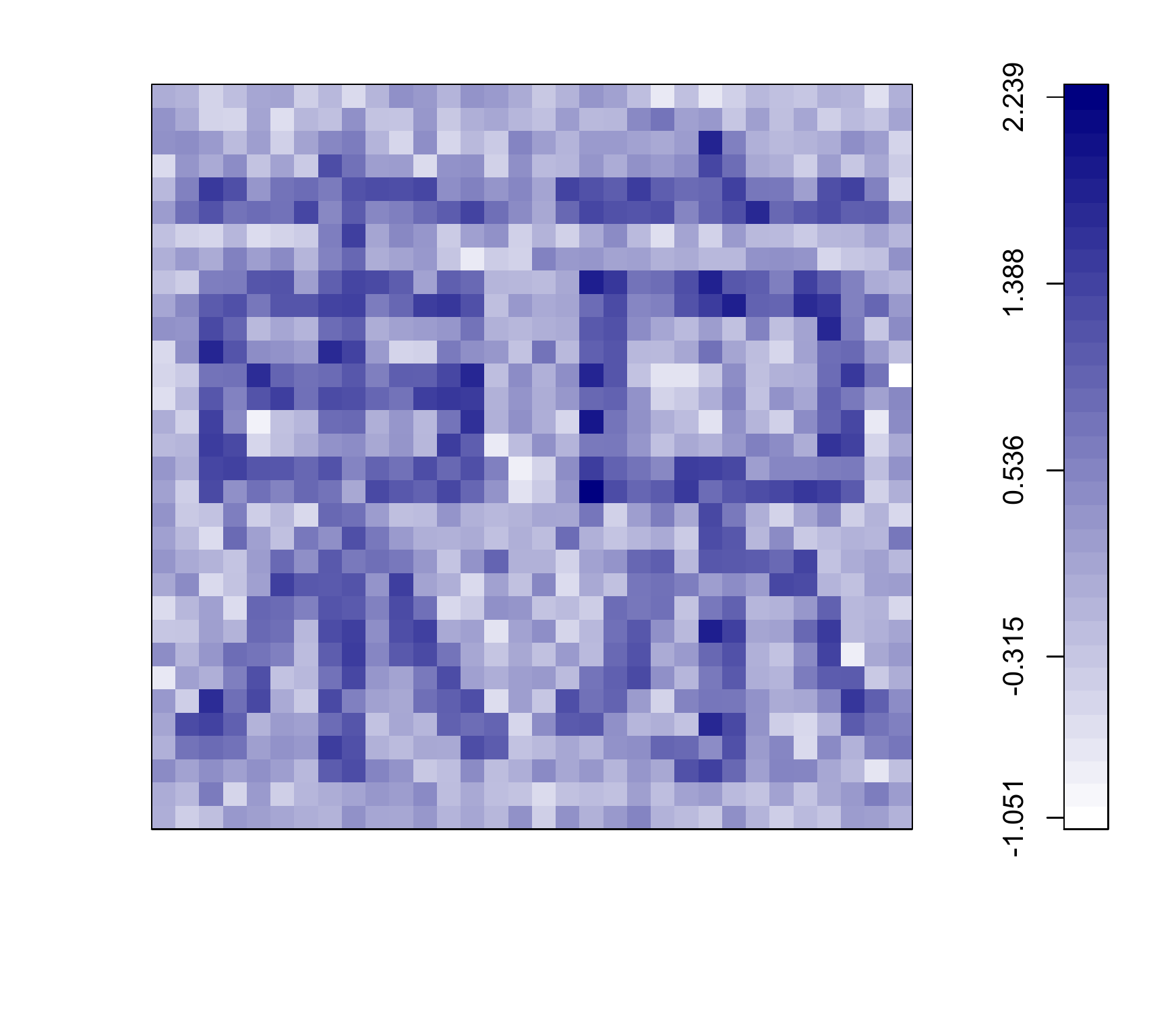}
 \end{minipage}
  \centering
 \begin{minipage}{0.49\columnwidth}
  \centering
   \includegraphics[bb=0 0 849 732 ,width=120mm,clip]{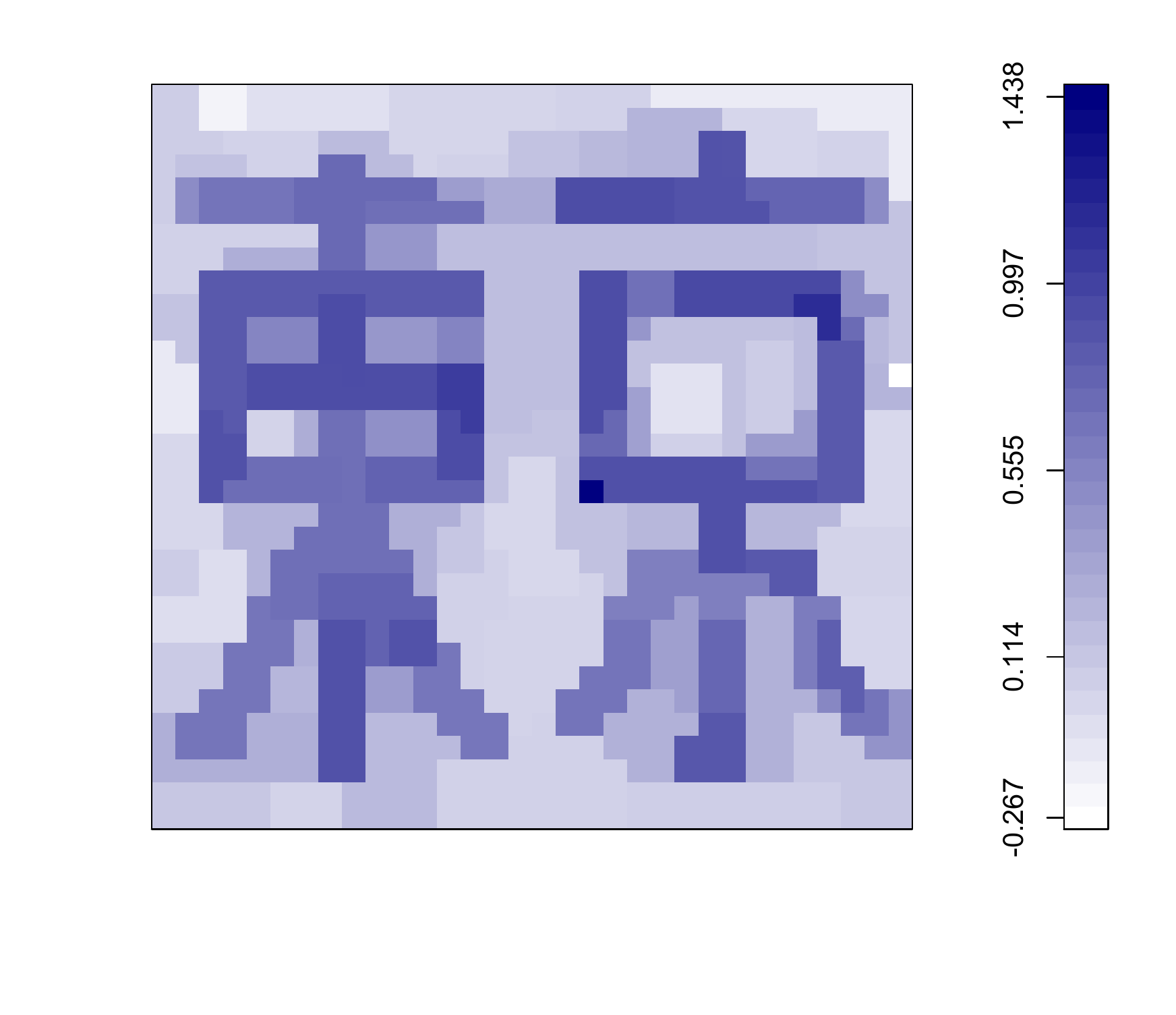}
 \end{minipage}
 \begin{minipage}{0.49\columnwidth}
  \centering
   \includegraphics[bb=0 0 849 732 ,width=120mm,clip]{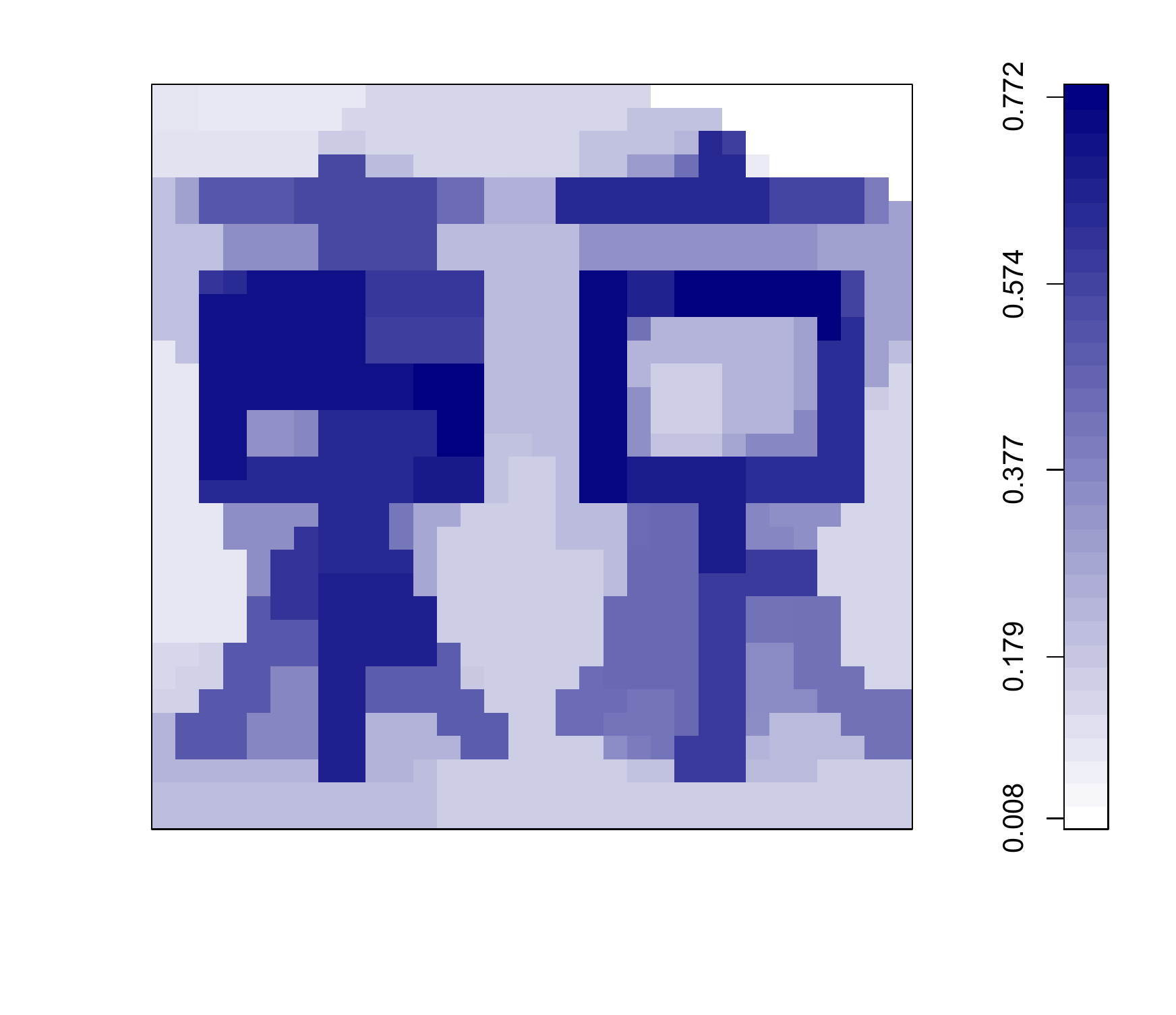}
 \end{minipage}
 \caption{
Results for artificial data generated from 2d fused lasso model.
Upper left panel: true image.
Upper right panel: noisy image.
Lower left panel: result of the proposed method.
Lower right panel: result of the non-Bayesian 2d fused lasso.
}
 \label{image_result}
\end{figure}

Next, we considered a numerical demonstration for the 2d fused lasso model applied to image reconstruction.
A sample image was generated by simulation. 
The upper left panel in Figure $\ref{image_result}$ shows the true image taking the values from $0$ (blue) to $1$ (white).
The upper right panel in Figure $\ref{image_result}$ shows a noisy image which has noises generated from normal distribution with mean $0$ and standard deviation $0.35$. 
These images are $32\times32=1024$ pixel in size. 
The hyper-parameters $(\lambda_2,\gamma_2)$ were tested for $(200,5)$ values and chosen by the EBIC. 
We compared the proposed method to the non-Bayesian 2d fused lasso by Friedman {\it et al}. (2007) which is implemented in the \texttt{genlasso} package in the software \texttt{R}. 
The regularization parameter was chosen by the EBIC.

The lower left and lower right panels in Figure $\ref{image_result}$ show respectively the results of the proposed method and those of the non-Bayesian 2d fused lasso.
The non-Bayesian 2d fused lasso failed to recognize a blue area in the true image as light blue. 
On the other hand, the proposed method correctly recognized the blue area in the true image blue. 
The result shows that the proposed method worked better than the non-Bayesian 2d fused lasso. 
The squares error $\| \vector{\beta}^*-\hat{\vector{\beta}} \|_2^2$ by the proposed method was $50.38$, while that by the non-Bayesian 2d fused lasso was $102.91$.
The results suggest that the proposed method may also be effective in image analysis.

%
%
%
%
%
%
%
%
%
%
%
\section{Concluding remarks}
%
%
%

We proposed the fused lasso-type estimation via NEG distribution for the penalty for differences between regression coefficients.
Because the NEG distribution has a more extreme spike at zero and more tail flatness than the Laplace distribution,
the proposed method enables us blocks to be estimated more clearly.
In addition, we proposed the sparse fused algorithm to provide a solution which has exactly zero coefficients and allows blocks to be estimated exactly.
Numerical examples showed that our proposed method provided a contrasted estimator, 
and worked better than existing methods.

It is important to extend the proposed method to other types of the generalized fused lasso method
as well as to develop information criteria such as the generalized Bayesian information criterion (GBIC; Konishi {\it et al.}, 2004) for evaluating these methods. 
We leave these interesting topics as future work.

%
%
%
\section*{Acknowledgments}
%
%
%

M. U. was supported by Grant-in-Aid for Young Scientist (B) (25870074) and Grants-in-Aid for Scientific Research (C) (25330049 and 25460403). 
S. K. was supported by Grant-in-Aid for Young Scientist (B) (15K15947). 
The computational resource was also provided by the Super Computer System, Human Genome Center, Institute of Medical Science, The University of Tokyo.

%
%
%
%
%
%
%
%
%
%


%
%
%
%
%
%
%

%

\begin{thebibliography}{30}
\bibitem{}
Akaike, H. (1973). Information theory and an extension of the maximum likelihood principle. {\it 2nd International Symposium on Information Theory} (Petrov, B.N. and Csaki, F., eds.), Akademiai Kiado, Budapest, pp. 267--281. (Reproduced in {\it Breakthroughs in Statistics}, Volume 1, S. Kotz and N. L. Johnson, eds., Springer Verlag, New York, (1992)).

\bibitem{}
Andrews, D. F. and Mallows, C. L. (1974). Scale Mixtures of Normal Distributions. {\it Journal of the Royal Statistical Society Series B}, {\bf 36}, 99--102.

\bibitem{}
Chen, J. and Chen, Z. (2008). Extended Bayesian information criterion for model
selection with large model space. {\it Biometrika}, {\bf 94}, 759--771.

\bibitem{}
Efron, B., Hastie, T., Johnstone, I. and Tibshirani, R. (2004). Least angle regression (with discussion). {\it Annals of Statistics}, {\bf 32}, 407--499.

\bibitem{}
Fan, J. and Li, R. (2001). Variable selection via nonconcave penalized likelihood and its oracle properties. {\it Journal of the American Statistical Association}, {\bf 96}, 1348--1360.

\bibitem{}
Friedman, J., Hastie, T., Hofling, H. and Tibshirani, R. (2007). Pathwise coordinate
optimization. {\it Annals of Applied Statistics}, {\bf 1}, 302--332.

\bibitem{} 
Griffin, J. and Brown, P. (2005). Alternative prior distributions for variable selection with very many more variables than observations. Technical report. University of Warwick, Coventry, UK.

\bibitem{} 
Hoerl, A. E. and Kennard, R. W. (1970). Ridge regression: biased estimation for nonorthogonal problem. {\it Technometrics}, {\bf 12}, 55--67. 

\bibitem{} 
Hoggart, C. J., Whittaker, J. C., De Iroio, M. and Balding, D. J. (2008). Simultaneous analysis of all SNPs in genome-wide and re-sequencing association studies. {\it PLOS Genetics}, {\bf 4}, e1000130. 


\bibitem{}
Jang, W., Lim, J., Lazar, N., Loh, J. and Yu, D. (2013). 
Regression shrinkage and grouping of highly correlated predictors with HORSES.
{arXiv:1302.0256}.




\bibitem{} 
Konishi, S., Ando, T. and Imoto, S. (2004). Bayesian information criteria and smoothing parameter selection in radial basis function networks. \textit{Biometrika}, {\bf 91}, 27--43. 

\bibitem{} 
Konishi, S. and Kitagawa, G. (2008). {\it Information Criteria and Statistical Modeling}. Springer, New York.

\bibitem{} 
Kyung, M., Gill, J., Ghosh, M. and Casalla, G. (2010). Penalized regression, standard error, and Bayesian lasso. {\it Bayesian Analysis}, {\bf 5}, 369--412.

\bibitem{} 
Park,\ T.\ and Casella,\ G.\ (2008).\ The Bayesian lasso.\ {\it Journal of the American Statistical Association},\ {\bf103},\ 681--686.

\bibitem{} 
Rockova, V. and Lesaffre, E. (2014).\ Incorporating grouping information in Bayesian variable selection with applications genomics. {\it Bayesian Analysis}, {\bf 9}, 221--258.

\bibitem{} 
Schwarz, G. (1978). Estimating the dimension of a model. {\it Annals of Statistics}, {\bf 6}, 461--464.

\bibitem{} 
Tibshirani,\ R.\ (1996).\ Regression shrinkage and selection via lasso.\ {\it Journal of the Royal Statistical Society Series B},\ {\bf58},\ 267--288.

\bibitem{} 
Tibshirani, R., Saunders, M., Rosset, S., Zhu, J. and Knight, K. (2005). Sparsity and
smoothness via the fused lasso. {\it Journal of the Royal Statistical Society Series B}, {\bf 67}, 91--108.

\bibitem{} 
Tibshirani, R. and Wang, P. (2008). Spatial smoothing and hot spot detection for CGH data using the fused
lasso. {\it Biostatistics}, {\bf 9}, 18--29.

\end{thebibliography}
\end{document}